\documentclass[11pt]{article}
\usepackage[english]{babel}
\usepackage[utf8]{inputenc}
\usepackage{lmodern}
\usepackage{array}
\usepackage[T1]{fontenc}
\usepackage[affil-it]{authblk}
\usepackage[toc,page]{appendix}
\usepackage[left=2.5cm,right=2.5cm,top=3cm,bottom=3cm]{geometry}
\usepackage{amsmath,amssymb,mathrsfs}
\usepackage{amsfonts}
\usepackage{graphicx}
\usepackage{tensor}
\usepackage[dvipsnames]{xcolor}
\usepackage[normalem]{ulem}

\usepackage[backend=bibtex,style=numeric-comp,maxnames=4,sorting=none,url=false,hyperref=true,eprint=true,doi=true]{biblatex}
\usepackage{hyperref}

\hypersetup{
	colorlinks=true,
	urlcolor=blue,
	linkcolor={black!40!black},
	citecolor={green!40!black}
}
\usepackage{mathtools}
\usepackage{cleveref}

\newcommand{\dd}{{\rm{d}}}        
\newcommand{\A}{{\mathcal{A}}}    
\newcommand{\B}{{\mathcal{B}}}    
\newcommand{\C}{{\mathcal{C}}}    
\newcommand{\D}{{\mathcal{D}}}    
\newcommand{\HH}{{\mathcal{H}}}   
\newcommand{\Z}{{\mathcal{Z}}}    
\renewcommand{\P}{\mathcal{P}}
\newcommand{\W}{\mathcal{W}}
\newcommand{\Q}{\mathcal{Q}}
\newcommand{\ct}[2]{\tensor{{#1}}{#2}} 
\newcommand{\ctr}[3]{\tensor[{#1}]{#2}{#3}}
\newcommand{\ctst}[2]{\tensor[_{{\,}_{\scriptscriptstyle\!\text{s}}\!}]{{#1}}{#2}}
\newcommand{\ctell}[2]{\tensor{{#1}}{#2}[\mathbf{l} ]}
\newcommand{\ctl}[2]{\tensor{{#1}}{#2}\big[{\ctst{\mathbf{l}}{}}\big]}
\newcommand{\ctlt}[2]{\tensor{{#1}}{#2}\big[\ctst{\tilde{\mathbf{{l}}}}{}\big]}
\newcommand{\ctkt}[2]{\tensor{{#1}}{#2}\big[\ctst{\tilde{\mathbf{{k}}}}{}\big]}
\newcommand{\ctk}[2]{\tensor{{#1}}{#2}\big[\ctst{\mathbf{k}}{}]}

\newcommand{\cd}[1]{\tensor{\nabla}{#1}}
\newcommand{\cds}[1]{\tensor{\overline{\nabla}}{#1}}

\newcommand{\cta}[2]{\tensor[_{{\,}_{\scriptscriptstyle\!\text{a}\!}\!}]{{#1}}{#2}}
\newcommand{\cts}[2]{\tensor{\overline{#1}}{#2}}

\def\scri{{\!\mathscr{J}}}
\def\scrisub{{_{\!\mathscr{J}}}}
\newcommand{\ms}[1]{\ct{h}{#1}}

\newcommand{\pt}[2]{\tensor{\hat{#1}}{#2}}

\newcommand{\defeq}{\vcentcolon=}
\newcommand{\defeqs}{\stackrel{\scri}{\vcentcolon=}}
\newcommand{\eqs}{\stackrel{\scri}{=}}
\newcommand{\prn}[1]{\left(#1\right)}

\newcommand{\brkt}[1]{\left[#1\right]}

\newcommand{\Bbrkt}[1]{\Bigg[#1\Bigg]}
\newcommand{\cbrkt}[1]{\left\lbrace#1\right\rbrace}

\def \bF {\mathbf{F}}
\def \bA {\mathbf{A}}

\newcommand{\rovno}{\!\!\!\!& = &\!\!\!\!}
\newcommand{\equi}{\!\!\!\!& \defeq &\!\!\!\!}
\newcommand{\ekviv}{\!\!\!\!& \equiv &\!\!\!\!}

\newcolumntype{M}[1]{>{\centering\arraybackslash}m{#1}}
\newcolumntype{N}{@{}m{0pt}@{}}
\usepackage{color}

\newcounter{mnotecount}[section]
\numberwithin{equation}{section}
\renewcommand{\themnotecount}{\thesection.\arabic{mnotecount}}
\newcommand{\mnote}[1]
{\protect{\stepcounter{mnotecount}}$^{\mbox{\footnotesize
$
\bullet$\themnotecount}}$ \marginpar{
\raggedright\tiny\em
$\!\!\!\!\!\!\,\bullet$\themnotecount: #1} }

\title{Analysis of gravitational radiation generated by type D black holes with positive cosmological constant}

\author[a]{Francisco Fernández-Álvarez \thanks{\tt francisco.fernandez@ehu.eus}}
\author[b]{\ Ji\v{r}\'{\i} Podolsk\'y \thanks{\tt jiri.podolsky@mff.cuni.cz}}
\author[a]{\ José M. M. Senovilla \thanks{\tt josemm.senovilla@ehu.eus}}
\affil[a]{Departamento de Física\protect\\ Universidad del País Vasco UPV/EHU\protect\\ Apartado 644, 48080 Bilbao, Spain\protect\\\ }
\affil[b]{Charles University, Faculty of Mathematics and Physics,\protect\\ Institute of Theoretical Physics, \protect\\ V~Hole\v{s}ovi\v{c}k\'ach 2, 18000 Prague 8, Czechia.}
\date{\today{}}
\begin{document}
\maketitle

\begin{abstract}
The criterion for existence of gravitational radiation at  conformal infinity in the presence of a positive cosmological constant is applied to a general family of exact solutions representing generic (pairs of) black holes of algebraic type~D. Our analysis shows that {\em only accelerating black holes generate gravitational radiation} measurable at infinity. This very satisfactory result confirms the goodness of the criterion. To that end, a new metric form of the family of exact type~D black holes is constructed  --- including any cosmological constant and a (double-aligned) non-null electromagnetic field --- whose expression is suitable for investigation of the asymptotic structure of this large family of spacetimes. The family depends on seven physical parameters, namely $m$, $a$, $l$, $\alpha$, $e$, $g$,  and $\Lambda$ that characterize mass, specific angular momentum parameter, NUT parameter, acceleration, electric and magnetic charges, and the cosmological constant, respectively.
\end{abstract}

\vfil\noindent
PACS class:  04.20.Jb, 04.70.Bw, 04.40.Nr, 04.70.Dy



\vfil
\eject
\tableofcontents
\section{Introduction}\label{sec:intro}
Gravitational waves and black holes are two of the most outstanding predictions of Einstein's General Relativity. There is a wide acceptance in the scientific community that these two features of Einstein's gravity theory match real physical phenomena in our universe. This consensus, laying on a solid theoretical background, is strongly supported by the first direct detection of a gravitational wave produced by a binary black hole merger, confirmed in 2016 \cite{Ligo-Virgo:2016}, and the observation of the shadow of supermassive black hole at the centre of M87, reported in 2019 \cite{EHT:2019}. These two recent discoveries, milestones of a new era of gravitational-wave astronomy and black hole imaging, provide observational data of physical processes under extreme conditions that are of great interest. On top of this, there is the third crucial observational evidence that traces back to the late nineties: the accelerated expansion of the universe \cite{Riess:1998,Perlmutter:1999} driven by a (maybe effective) positive cosmological constant ${\Lambda>0}$. \\

For over a century now, the theory has developed a solid description of the gravitational-radiation physics.
In the full non-linear regime, much progress crystallized in the late fifties and sixties \cite{Pirani:1957,Bel:1958,Bondi:1962,Sachs:1962,Penrose:1962}, leading to a rigorous formulation that served to understand gravitational radiation in asymptotically flat regions of a spacetime, bringing into a robust mathematical basis \cite{Geroch:1977} concepts such as the energy emitted by generic sources via gravitational waves, or the characterization of its existence (at infinity) based on the well-known News criterion.  Unfortunately, this mathematical formulation only applied to the case  ${\Lambda=0}$, thus excluding the description of gravitational radiation and the asymptotic structure of spacetimes with ${\Lambda>0}$, which remained an open problem  \cite{Penrose-Rindler:1985bww,Penrose:2011} until recently --- last decade ---, when new advances in that direction appeared \cite{Ashtekar:2014,Szabados:2015,Fernandez-Senovilla:2020,KrtousPodolsky:2004}.

In particular, now there is a general framework developed in \cite{Fernandez-Senovilla:2022b} to study the asymptotic structure of spacetimes with a positive $ \Lambda $ --- see \cite{Senovilla:2022} for a review of this work, references to other approaches  and the relation to the traditional case with ${\Lambda=0}$. This framework provides a neat criterion, based on the commutator of the {\em canonical electric and magnetic parts} of the re-scaled {\em Weyl tensor at conformal infinity}~$\scri$, for the existence or absence of gravitational radiation escaping from the spacetime (reaching $\scri^+$). This criterion was tested on several examples to lead to correct results, and to reduce to the traditional News criterion in the case of vanishing $\Lambda$ \footnote{Apart from that, there is further work to be done as application of this framework. For example, to test the quadrupole formulae with $ \Lambda>0 $, memory effects and other works carried out in the \emph{linearised regime} \cite{He2017,Kolanowski:2020,Compere:2023}, to see the connection with different notions of mass such as the one in  \cite{Szabados:2015}, or to test the Bondi-like form of the metric with $ \Lambda>0 $ \cite{Bonga:2023}. All these applications are of high interest too, but they are out of the scope of the present work.}. \\

This work encompasses the investigation of the afore-mentioned two aspects of the theory: gravitational radiation and black-hole exact solutions with a positive cosmological constant ${\Lambda>0}$. Our goal here is to further test the radiation criterion developed in \cite{Fernandez-Senovilla:2022b}  in a sufficiently general class of exact black-hole solutions. Actually, the black-hole fauna has been getting larger recently with new possibilities \cite{ChngMannStelea:2006gh,Podolsky:2020xkf,AstorinoBoldi:2023,Astorino:2023,Astorino:2024,Astorino:arXiv}. Let us recall, however,  that the simplest (and first) identification of gravitational radiation for black-hole spacetimes  with $\Lambda$ was obtained for the accelerated solution called C-metric \cite{Ashtekar-Dray:1981,KrtousPodolsky:2003tc,PodolskyOrtaggioKrtous:2003gm,Podolsky-Kadlecova:2009,Fernandez-Senovilla:2020}, extending previous studies of radiation in the C-metric in asymptotically flat settings. This solution is included in the wider class of type~D exact solutions found by Plebański and Demiański \cite{Plebanski-Demianski:1976, Griffiths-Podolsky:2009} --- see  \cite{Stephani-etal:2003, Griffiths-Podolsky:2006} for review and further references, and thus this wider class is a perfect arena to put our ideas to test.

To that end, a new form presented in \cite{Podolsky-Vratny:2021, Podolsky-Vratny:2023} will be further improved herein, so that it is amenable to a proper analysis of future infinity $\scri^+$. This new form may be of interest on its own. The conclusion that we reach is very satisfactory: {\em gravitational radiation exists only if the black holes are accelerating}.

The case with ${\Lambda=0}$ can be equally studied, and is actually implicitly included, by taking the appropriate limit \cite{Fernandez-Senovilla:2022a},  while the case ${\Lambda<0}$ is postponed for later investigation.

\subsection*{Conventions and notation}
The conventions used throughout this paper are the same ones as employed in \cite{Fernandez-Senovilla:2022b}, together with some notation from \cite{Podolsky-Vratny:2023,KrtousPodolsky:2004}.   For reader's convenience, the key symbols and some definitions are summarised in \cref{tab:conventions}.\\

		\begin{table}[h!]
			\centering
			\begin{tabular}{|M{5cm}| M{3cm} | M{3cm} |M{3cm} |N}
				\hline
				\quad&  Physical spacetime $ \hat{M} $& Conformal spacetime $ {M} $ &  Conformal infinity $ \scri $ ($ \Lambda>0 $) &\\ \hline
				\vspace{0.2cm}\textbf{Metric}& $ \ct{\hat{g}}{_{\alpha\beta}} $ & $ \ct{g}{_{\alpha\beta}}  $ & $ \ms{_{ab}} $   &\\[0.2cm] \hline
					\vspace{0.2cm}\textbf{Volume form}&	$ \pt{\eta}{_{\alpha\beta\gamma\delta}} $& $ \ct{\eta}{_{\alpha\beta\gamma\delta}} $  & $ \ct{\epsilon}{_{abc}} $  &  \\[0.2cm]\hline
				\textbf{Covariant derivative}& $ \hat{\nabla}{_{\alpha}} $	&$ \cd{_{\alpha}} $	&	$ \cds{_{a}} $&  \\[0.2cm]\hline		 \vspace{0.2cm}			 
				\textbf{Riemann tensor}& $ \pt{R}{_{\alpha\beta\gamma}^\delta} $ 	& $ \ct{R}{_{\alpha\beta\gamma}^\delta} $ & $ \cts{R}{_{abc}^d} $  & \\[0.2cm] \hline	\vspace{0.2cm}
				\textbf{Weyl tensor}& $ \pt{C}{_{\alpha\beta\gamma}^\delta} $ ($=\ct{C}{_{\alpha\beta\gamma}^\delta}   $)	& $ \ct{C}{_{\alpha\beta\gamma}^\delta} =\Omega\, \ct{d}{_{\alpha\beta\gamma}^\delta}$ & --  &  \\[0.2cm]\hline	\vspace{0.2cm}
				\textbf{Projector}& --	& -- & $ \ct{P}{^\alpha_\beta} $  &  \\[0.2cm]\hline	 \vspace{0.2cm}
				\textbf{Orthonormal tetrad}&	$ {(\hat{\mathbf{t}},\hat{\mathbf{q}},\hat{\mathbf{r}},\hat{\mathbf{s}})} $ & $ {({\mathbf{t}},{\mathbf{q}},{\mathbf{r}},{\mathbf{s}})} $	&--&  \\[0.2cm]\hline	\vspace{0.2cm}
				\textbf{Null complex tetrad}&	$ {(\hat{\mathbf{k}},\hat{\mathbf{l}},\hat{\mathbf{m}})} $ & $ {(\mathbf{k},\mathbf{l},\mathbf{m})} $	 & --&  \\[0.2cm]\hline	
			\end{tabular}
			\caption{Some key symbols used in the paper. See the bullet points in the text for related definitions.}\label{tab:conventions}
		\end{table}

\textbf{Signature, indices and curvature}
		\begin{itemize}
				\item Spacetime metric signature: $ \prn{-,+,+,+} $.
				\item Spacetime indices: $ \alpha,\beta,\gamma,$ etc; three-dimensional space-like hypersurfaces indices: $ a,b,c,$ etc.
				\item Riemann tensor, Ricci tensor, and scalar curvature: $ \ct{R}{_{\alpha\beta\gamma}^\delta}\,\ct{v}{_\delta}\defeq\prn{\cd{_{\alpha}}\cd{_\beta}-\cd{_{\beta}}\cd{_{\alpha}}}\ct{v}{_\gamma} $, $ \ct{R}{_{\alpha\beta}}\defeq\ct{R}{_{\alpha\mu\beta}^\mu} $, $ R\defeq\ct{R}{_{\mu\nu}}\,\ct{g}{^{\mu\nu}} $.
				\item Orientation: in an orthonormal basis $ \ct{\eta}{_{0123}}=1 $, $ \ct{\epsilon}{_{123}}=1 $; in a null basis $ \ct{\eta}{_{\hat{0}\hat{1}\hat{2}\hat{3}}}= \mathrm{i}\,  $.
			\end{itemize}

\textbf{Relevant fields and decorations}
			\begin{itemize}
			\item Conformal factor $ \Omega $.
			Conformal infinity ${\scri:=\{\Omega =0\}}$.
			Conformal gauge-transformed quantities $ T\rightarrow\tilde{T} $.
			\item In general, quantities in physical spacetime ${(\hat{M},\pt{g}{_{\alpha\beta}})} $ carry a hat ($ \pt{T}{} $), to distinguish them from those in conformal (unphysical) spacetime $ {(M,\ct{g}{_{\alpha\beta}})} $.
			\item Conformal and physical Weyl tensor are ${\ct{C}{_{\alpha\beta\gamma}^\delta} =\pt{C}{_{\alpha\beta\gamma}^\delta}}$, and the rescaled version is
			$$ \ct{d}{_{\alpha\beta\gamma}^\delta}\defeq \Omega^{-1}\,\ct{C}{_{\alpha\beta\gamma}^\delta} \ .$$
			\item Normal to the hypersurfaces $\Omega =$ const., in particular to $\scri$,is ${{\ct{N}{_{\alpha}}\defeq \cd{_{\alpha }}\Omega}}$. Unit normal to those hypersurfaces and to $ \scri $ are ${\sqrt{-\ct{N}{^{\mu}}\ct{N}{_{\mu}}}\,\ct{n}{^\alpha}\defeq \ct{N}{^\alpha} }$.\footnote{This is well defined at $ \scri $ and in its neighborhood, as it is assumed that ${\Lambda>0}$. See the main text for more details on this.}
			\item  Projector to $ \scri $: $ \ct{P}{^\alpha_{\beta}}=\delta^\alpha_{\,\,\beta}+\ct{n}{^\alpha}\ct{n}{_{\beta}} |_\scri$.
			\item Basis of vector fields tangent to $ \scri $: $ \{\ct{e}{^{\alpha}_{a}} \}$\,. They can be used to pullback tensors to $\scri$.
			\item Electric $ \ct{D}{_{\alpha\beta}}\defeq \ct{d}{_{\alpha\mu\beta\nu}}\,\ct{n}{^\mu}\ct{n}{^\nu}$ and magnetic $ 2\,\ct{C}{_{\alpha\beta}}\defeq  2\,\ctr{^*}{d}{_{\alpha\mu\beta\nu}}\,\ct{n}{^\mu}\ct{n}{^\nu} =\ct{d}{_{\rho\sigma\beta\nu}}\,\ct{\eta}{^{\rho\sigma}_{\alpha\mu}}\,\ct{n}{^\mu}\ct{n}{^\nu} $ parts of the rescaled Weyl tensor $ \ct{d}{_{\alpha\beta\gamma}^\delta} $ with respect to the unit, timelike $ \ct{n}{^\alpha} $. Their pull-backed versions, defined only on $\scri$, are
			$$
			D_{ab} = D_{\alpha\beta}\,e^\alpha{}_a\, e^\beta{}_b \,, \hspace{5mm}
            C_{ab} = C_{\alpha\beta}\,e^\alpha{}_a\, e^\beta{}_b \,.
			$$
			\end{itemize}
			
\section{New form of the metric and conformal compactification}\label{sec:newformmetric}
In \cite{Podolsky-Vratny:2021, Podolsky-Vratny:2023}, a convenient representation of the \emph{family of type D black holes} was found. The metric --- recalled in  \cref{app:derivation-metric} as eqs.~\eqref{newmetricGP2005}--\eqref{finalQagain} --- nicely represents this large class of exact spacetimes.\footnote{ \label{foot2} Except that it fails to include a special  degenerate subcase of {\em accelerating purely NUT} black holes because vanishing spin parameter (${a=0}$) implies vanishing acceleration (${\alpha=0}$)  of the black hole. An improved set of physical parameters which circumvents this problem has been recently found by Astorino \cite{Astorino:2024,Astorino:arXiv}.} Moreover, it naturally generalizes the standard forms of famous (rotating, charged, accelerating) black hole solutions, with two black-hole horizons (outer and inner) and two cosmological/acceleration horizons, namely the Kerr-Newman-(A)dS black holes, charged Taub-NUT-(A)dS, their accelerated versions, and other black holes. These can immediately be obtained as direct subcases by setting the corresponding physical parameter(s) to zero. As shown in \cite{Podolsky-Vratny:2021, Podolsky-Vratny:2023}, various physical and geometrical properties can more easily be studied, such as their singularities, horizons, ergoregions, global structure, cosmic strings, or thermodynamics.

However, investigation of the \emph{asymptotic structure} of this large family of important spacetimes is quite involved. The global extension across all horizons can be performed, and the Penrose conformal diagrams can be constructed by employing a sophisticated set of transformations. But the \emph{conformal infinity}, whose character is crucial for determining the \emph{radiative} properties of these black holes, is not readily identified even in the improved form of the metric \eqref{newmetricGP2005}--\eqref{finalQagain}. Here, an alternative metric which overcomes this problem is presented --- for its derivation from the original one and further details see \cref{app:derivation-metric}. In coordinates $\cbrkt{t,q,\theta,\varphi}$ with ranges $ t\in\prn{-\infty,\infty} $,  $ q\in \prn{-\infty,\infty} $, 
$ \theta\in\brkt{0,\pi}  $ and $ \varphi\in\left[0,2\pi C\right) $,\footnote{The range of $q$ is as shown for the general case with $l\neq 0$. However, for some special cases with $l=0$ the range might have to be restricted to $q\in (0,\infty)$. On the other hand, the parameter $ C $ is related to the deficit angle along the axes, or conicity. See \cite{Griffiths-Podolsky:2009,Podolsky-Vratny:2023} for further details on this.} this new form of the metric reads
\begin{align}
\dd \hat{s}^2 = \frac{1}{\Omega^2} &
  \bigg(\!-\frac{Q}{\rho^2}\Big[\,\dd t- \big(a\sin^2\theta +4l\sin^2\!\tfrac{1}{2}\theta \big)\dd\varphi \Big]^2
   + \frac{\rho^2}{Q}\,\dd q^2  \nonumber\\
& \quad  + \,\frac{\rho^2}{P}\,\dd\theta^2
  + \frac{P}{\rho^2}\,\sin^2\theta\, \Big[\, a\,q^2\,\dd t -\big(1+(a+l)^2q^2\big)\,\dd\varphi \Big]^2
 \bigg), \label{newmetricJP2023}
\end{align}
where
\begin{eqnarray}
\Omega(q,\theta)    \rovno q - \frac{\alpha\,a}{a^2+l^2}\, (l+a \cos \theta) \,, \label{newOmega-q}\\[2mm]
\rho^2(q,\theta)    \rovno 1 + q^2\,(l+a \cos \theta)^2 \,, \label{newrho-q}
\end{eqnarray}
and
\begin{eqnarray}
P(\theta) \rovno 1 - 2\,\Big(\,\frac{\alpha\,a}{a^2+l^2}\,\,m - \frac{\Lambda}{3}\,l \Big)(l+a\,\cos\theta)\nonumber\\
      &&\hspace{-2mm}  +\Big(\frac{\alpha^2 a^2}{(a^2+l^2)^2} (a^2-l^2 + e^2 + g^2) + \frac{\Lambda}{3} \Big)(l+a\,\cos\theta)^2 \,,
 \label{newP-q}\\[2mm]
Q(q) \rovno \Big[\,1 - 2m\, q  + (a^2-l^2+e^2+g^2)\,q^2 \Big]
            \Big( q+\alpha\,a\,\frac{a-l}{a^2+l^2} \Big)
            \Big( q-\alpha\,a\,\frac{a+l}{a^2+l^2} \Big)\nonumber\\
   &&\hspace{-2mm}  - \frac{\Lambda}{3} \Big[\,1 + 2\alpha\,a\,l\,\frac{a^2-l^2}{a^2+l^2}\,q + (a^2+3l^2)\,q^2\,\Big]. \label{newQ-q}
\end{eqnarray}

There are two independent Killing vector fields, $ \partial_{\varphi} $ and $ \partial_{t} $. The function $ Q(q) $ has up to 4 different roots which give the black-hole and cosmo-acceleration horizons \cite{Podolsky-Vratny:2023}. They happen to be Killing horizons for a particular Killing vector. The spacetime depends on \emph{seven physical parameters}, whose interpretation is \cite{Podolsky-Vratny:2021, Podolsky-Vratny:2023}
\begin{align}
m \quad.....\quad & \hbox{mass parameter}     \nonumber \,,\\
a \quad.....\quad & \hbox{spin parameter (or Kerr-like rotation)} \nonumber \,,\\
l \quad.....\quad & \hbox{NUT parameter}      \nonumber \,,\\
e \quad.....\quad & \hbox{electric charge}    \nonumber \,,\\
g \quad.....\quad & \hbox{magnetic charge}    \nonumber \,,\\
\alpha \quad.....\quad & \hbox{acceleration parameter}  \nonumber \,,\\
\Lambda\quad.....\quad & \hbox{cosmological constant}  \nonumber \,.
\end{align}

For ${e,g}$ non-zero the black holes are charged, with their electromagnetic field represented by the 1-form potential
\begin{equation}
 \bA =  - \frac{\,e\,q + g\,(l+a \cos \theta)\,q^2}{1+(l+a \cos \theta)^2 \,q^2}\, \dd t
 + \frac{(e\,q + g\,l\,q^2) R + g\,S \cos \theta}{1+(l+a \cos \theta)^2 \,q^2} \,\dd\varphi\,,
\label{vector-potential-repeated}
\end{equation}
see (\ref{vector-potential}), (\ref{Fab}), where the auxiliary functions are
\begin{eqnarray}
R(\theta)  \rovno a\sin^2\theta + 2 l (1-\cos\theta) \,, \label{def-R}\\[2mm]
S(q)       \rovno 1 + (a+l)^2 q^2   \,. \label{def-S}
\end{eqnarray}
Interestingly,
	\begin{equation} \label{S-aq^2R}
			\rho^2 = S - a\,q^2 R \,.
	\end{equation}

Using the factorized  forms (\ref{newP}), (\ref{newQ}) of the metric functions,  alternative expressions for $ P $ and $ Q $ are
\begin{eqnarray}
P(\theta) \rovno \Big(1 -\frac{\alpha\,a}{a^2+l^2}\,r_{\Lambda+}\,(l+a\,\cos\theta) \Big)
            \Big(1 -\frac{\alpha\,a}{a^2+l^2}\,r_{\Lambda-}\,(l+a\,\cos\theta) \Big) , \label{newP-q-fact}\\[1mm]
Q(q) \rovno \big(1 -r_{\Lambda+}\,q \big) \big( 1-r_{\Lambda-}\,q \big)
            \Big(q +\alpha\,a\,\frac{a-l}{a^2+l^2}\Big)
            \Big(q -\alpha\,a\,\frac{a+l}{a^2+l^2}\Big) - \frac{\Lambda}{3}\,\Big[\,1 + \frac{(a^2+l^2)^2}{\alpha^2 a^2}\,q^4\,\Big],\qquad \label{newQ-q-fact}
\end{eqnarray}
 where the constants $r_{\Lambda\pm}$ are defined in \eqref{r+}.

Let us remark again that, due to the form of our metric, the vanishing of $a$ implies the absence of $\alpha$ --- except when one sets $l=0$ first, leading to the accelerating Kerr-Newman-(A)dS black holes and its subcases (see Appendix \ref{A1}), in which the parameter-space degeneracy disappears. The generic parameter-space degeneracy, which prevents us to identify the accelerating purely NUT black holes, is improved in a recent new parametrization and coordinate representation of the complete family of type~D black holes due to Astorino \cite{Astorino:2024,Astorino:arXiv}.\\

Now, the corresponding conformal metric $\ct{g}{_{\alpha\beta}}$ can readily be obtained from \eqref{newmetricJP2023}.

\subsection{Conformal metric and  conformal boundary $\scri$}
The next task is to explicitly derive the corresponding \emph{ metric $h_{ab}$ on the conformal boundary $ \scri $}. First, recall that the physical metric \eqref{newmetricJP2023} is related to the conformal one through the conformal factor ${\Omega>0}$,
	\begin{equation}
	g_{\alpha\beta} \defeq \Omega^2\,\hat{g}_{\alpha\beta}\,,
	\label{metrics}
	\end{equation}
	so that the unphysical line-element reads
	\begin{align}
	\dd {s}^2 = &
	  - \frac{Q}{\rho^2}\big[\,\dd t- R\,\dd\varphi \big]^2  + \frac{P}{\rho^2}\,\sin^2\theta\, \big[ a\,q^2\,\dd t - S\,\dd\varphi \big]^2
      + \frac{\rho^2}{Q}\,\dd q^2 + \,\frac{\rho^2}{P}\,\dd\theta^2 \,. \label{conformal-metricJP2023}
	\end{align}

Conformal infinity $ \scri $ is located at
	\begin{equation}
	  \Omega = 0\,.
	  \label{Omega=0}
	\end{equation}
	In view of (\ref{newOmega-q}) this is now simply given by the condition
	\begin{equation}
	 \scri\,:\qquad    q = \A + \B \cos \theta\,,
	  \label{scri-AB}
	\end{equation}
	where the convenient combinations of the three physical parameters are
	\begin{equation}
	 \A \defeq  \frac{\alpha\,a\,l}{a^2+l^2} \qquad\hbox{and}\qquad
	 \B \defeq  \frac{\alpha\,a^2}{a^2+l^2}\,,
	  \label{AB}
	\end{equation}
	implying the relations
	\begin{equation}
	 \A+\B = \frac{\alpha\,a}{a^2+l^2}\,(a+l) \qquad\hbox{and}\qquad
	 a\,\A = l\, \B \,.
	  \label{A+B}
	\end{equation}
	For \emph{any} angular coordinate ${\theta\in[0,\pi]}$, the value \eqref{scri-AB} of the coordinate $q$ corresponding to  $\scri$ is thus \emph{finite}. This circumvents the problem with the original form of the conformal factor (\ref{newOmega}). We also immediately see that whenever ${\alpha\,a=0}$, i.e. for black holes \emph{without acceleration} (${\alpha=0}$) or \emph{without the Kerr-like rotation} (${a=0}$), the conformal infinity $ \scri $ is simply located at
	\begin{equation}
	 \scri\,:\qquad     q = 0\,,
	  \label{scri-alpha-a=0}
	\end{equation}
	or, in the reciprocal coordinate ${r \defeq 1/q}$, at ${r=\infty}$. This is, of course, the expected result for the Kerr-Newman-(anti-)de Sitter  and also NUT-(anti-)de Sitter black holes.\\
	
	Evaluation on $ \scri $ of the metric functions (\ref{newrho-q}) and (\ref{newQ-q}), using the condition (\ref{scri-AB}), yields compact explicit expressions (we use a subscript $\scri$ to denote the restriction of any object to $\scri$)
	\begin{eqnarray}
	\rho^2_\scrisub (\theta)  \rovno  1 + \frac{(a^2+l^2)^2}{\alpha^2 a^2}\,(\A+ \B \cos \theta)^4 \nonumber\\
	   \rovno 1 + \frac{\alpha^2 a^2}{(a^2+l^2)^2}\,(l+a \cos \theta)^4\,, \label{rho-on-scri}\\[2mm]
	Q_\scrisub (\theta)       \rovno -\frac{\Lambda}{3}\,\rho^2_\scrisub (\theta) - \frac{\alpha^2 a^4}{(a^2+l^2)^2}
	   \,P(\theta) \,\sin^2\theta  \nonumber\\
	   \rovno -\frac{\Lambda}{3} - \frac{\alpha^2 a^2}{(a^2+l^2)^2}\,
	   \bigg[ \,\frac{\Lambda}{3}\, (l+a \cos \theta)^4 + a^2 P(\theta)\, \sin^2\theta \bigg]
	\,. \label{Q-on-scri}
	\end{eqnarray}
	Because $P(\theta)$ is independent of $q$, it remains the same on $ \scri $,
	\begin{eqnarray}
	P_\scrisub(\theta)  \defeq  P(\theta)  \,. \label{P-on-scri}
	\end{eqnarray}
	It is given by \eqref{newP-q}, or by  its special factorized form \eqref{newP-q-fact}, that is
	\begin{eqnarray}
	P (\theta) \rovno \Big(1 - r_{\Lambda+} (\A+\B \cos\theta) \Big)
	                  \Big(1 - r_{\Lambda-} (\A+\B \cos\theta) \Big) \,. \label{newP-q-fact/compact}
	\end{eqnarray}

	The important relation (\ref{Q-on-scri}) can be rewritten as
	\begin{eqnarray}
	\big(Q + \B^2P\sin^2\theta\big)_\scrisub \rovno -\frac{\Lambda}{3}\,\rho^2_\scrisub    \,, \label{Q-on-scri-short}
	\end{eqnarray}
	which enables us to simplify the last two terms in the conformal metric (\ref{conformal-metricJP2023}) to the form
	\begin{align}
	\Big( \frac{\rho^2}{Q}\,\dd q^2  + \,\frac{\rho^2}{P}\,\dd\theta^2 \Big)_\scrisub =
	   -\frac{\Lambda}{3}\,\Big( \frac{\rho^4}{PQ}\Big)_\scrisub\,\dd\theta^2 \,. \label{conformal-metric-simplification-1}
	\end{align}

	In order to do the pullback to $ \scri $, and obtain the induced metric there, this must be combined with the first terms containing $\dd t$ and $\dd\varphi$ in (\ref{conformal-metricJP2023}). Using (\ref{Q-on-scri-short}) and (\ref{scri-AB}), they take the form
	\begin{align}
	&
	  \frac{\Lambda}{3}\,\big[\,\dd t- \big(a\sin^2\theta +4l\sin^2\!\tfrac{1}{2}\theta \big)\dd\varphi \big]^2
	  \ + \ \frac{P}{\rho^2_\scrisub}\sin^2\theta\>\times
	  \label{conformal-metric-simplification-2}\\
	& \bigg[\B^2\big[\,\dd t- \big(a\sin^2\theta +4l\sin^2\!\tfrac{1}{2}\theta \big)\dd\varphi \big]^2
	  + \big[ a\,(\A + \B \cos \theta)^2\,\dd t -\big(1+(a+l)^2(\A + \B \cos \theta)^2\big)\,\dd\varphi \big]^2 \bigg].
	  \nonumber
	\end{align}
	Employing the relations (\ref{A+B}) and the identities
	\begin{eqnarray}
	a\sin^2\theta +4l\sin^2\!\tfrac{1}{2}\theta  \ekviv (1-\cos\theta)\,\big[ (a+l) + (l+a \cos \theta) \big]
   \equiv R \,, \label{ident-1}\\[1mm]
	\B\,\big[ (a+l) + (l+a \cos \theta) \big]    \ekviv a\,\big[ \,(\A+\B) + (\A + \B \cos \theta) \big]  \,, \label{ident-2}\\[1mm]
	a^2 (1-\cos\theta)^2    \ekviv \frac{(a^2+l^2)^2}{\alpha^2 a^2}\,\big[ \,(\A+\B) - (\A + \B \cos \theta) \big]^2  \,, \label{ident-2}\\[1mm]
	(\A + \B)^2 + (a+l)^2(\A+ \B \cos \theta)^4  \ekviv \alpha\,\B\,\frac{(a+l)^2}{a^2+l^2}\,\rho^2_\scrisub  \,, \label{ident-3}
	\end{eqnarray}
	the big square bracket in the second line of (\ref{conformal-metric-simplification-2}) factorises and simplifies considerably to
	\begin{align}
	& \rho^2_\scrisub \bigg[\,\B^2\dd t^2
	  - 2\, \B\, \alpha a \frac{(a+l)^2}{a^2+l^2}\,\dd t\,  \dd \varphi\
	  + \C^2 \,\dd\varphi^2 \bigg]\,,
	\end{align}
	where
	\begin{align}
	\C^2  = \frac{1}{\rho^2_\scrisub}\,\bigg[ \B^2 (1-\cos\theta)^2\big[ (a+l) + (l+a \cos \theta) \big]^2
	+ \big[ 1+(a+l)^2(\A + \B \cos \theta)^2\, \big]^2
	\bigg] \, ,
	\end{align}
	or equivalently
	\begin{align}
	\C^2  =
	\frac{1}{\rho^2_\scrisub}\,\bigg[ a^2 (1-\cos\theta)^2  \big[  \,(\A+\B) + (\A + \B \cos \theta) \big]^2
	+ \big[ 1+(a+l)^2(\A + \B \cos \theta)^2\, \big]^2 \bigg].
	\end{align}
	Quite surprisingly, this complicated-looking function can be simplified to the following \emph{constant}
	\begin{align}\label{defC}
	\C^2=1 + \alpha^2 a^2 \frac{(a+l)^4}{(a^2+l^2)^2} \,.
	\end{align}

	Putting all the terms of \eqref{conformal-metricJP2023} together, the {\em conformal metric $h_{ab}$ on  the $\Lambda>0$ $\scri$  is}
	\begin{align}
	h =&\   \frac{\Lambda}{3}\,\Big[\,\dd t- \big(a\sin^2\theta +2l(1-\cos\theta) \big)\dd\varphi \Big]^2
	    \nonumber\\
	&  \ + P \sin^2\theta\,\bigg[\,\B^2\dd t^2
	  - 2\, \B\, \alpha a \frac{(a+l)^2}{a^2+l^2}\,\dd t\,  \dd \varphi\
	  + \C^2 \,\dd\varphi^2 \bigg]
	-\frac{\Lambda}{3}\,\frac{\rho_\scrisub^4}{P\,Q_\scrisub}\,\dd\theta^2 \, .
	    \label{conformal-metric-final}
	\end{align}
	The metric coefficients are thus
	\begin{align}
	h_{tt} & =  \frac{\Lambda}{3} + \B^2 P\sin^2\theta \, , \nonumber\\
	h_{t\varphi} & = - \frac{\Lambda}{3}\,R
	 - \alpha a\, \frac{(a+l)^2}{a^2+l^2} \,\B\, P \sin^2\theta \,,
	    \nonumber\\
	h_{\varphi\varphi} & = \frac{\Lambda}{3}\,R^2
	 +\C^2 P \sin^2\theta \,, \label{h-metric-coefficients}\\
	h_{\theta\theta} & = -\frac{\Lambda}{3}\,\frac{\rho_\scrisub^4}{P\,Q_\scrisub} \,, \nonumber
	\end{align}
	where $P$ is given by  \eqref{newP-q} or in the factorized form \eqref{newP-q-fact/compact}, $\rho_\scrisub$ is given by \eqref{rho-on-scri}, $Q_\scrisub$ is given by \eqref{Q-on-scri}, and $R$ is defined in \eqref{def-R}. The constants $\B$ and  $\C$ are determined by \eqref{AB} and \eqref{defC}, respectively. Therefore, the metric \eqref{conformal-metric-final} can be explicitly written as
	\begin{align}
	h =&\   \frac{\Lambda}{3}\,\Big[\,\dd t- \big(a\sin^2\theta +2l(1-\cos\theta) \big)\dd\varphi \Big]^2
	    \nonumber\\
	&  \ + P \sin^2\theta\,\bigg[\,\dd\varphi^2 +
	 \frac{\alpha^2 a^2}{(a^2+l^2)^2} \big( a\,\dd t - (a+l)^2\,\dd\varphi \big)^2 \bigg]
	-\frac{\Lambda}{3}\,\frac{\rho_\scrisub^4}{P\,Q_\scrisub}\,\dd\theta^2 \, .
	    \label{conformal-metric-final-alternative}
	\end{align}
This metric is \emph{everywhere regular}, except on the intersection with the horizons (given by ${Q=0}$) and at the axes (given by ${\sin\theta=0}$, that is at ${\theta=0}$ and ${\theta=\pi}$).\\

The metric \eqref{conformal-metric-final-alternative} can also be expressed in the following convenient form
	\begin{align}
	h = \omega^2 \Big[ \big( \dd t +A\, \dd \varphi \big)^2 +W^2 \dd\varphi^2 +H^2 \dd\theta^2 \Big],
	\label{conformal-metric-final-alternative-1}
	\end{align}
	with
	\begin{align}
	\omega^2 & \defeq  \frac{\Lambda}{3} +\B^2P\sin^2\theta,\\
	A   & \defeq  - \omega^{-2}\, \bigg[ \frac{\Lambda}{3}\,R
         +\alpha a \frac{(a+l)^2}{a^2+l^2} \,\B\, P \sin^2\theta \bigg],\\
	H^2 & \defeq  - \omega^{-2}\, \frac{\Lambda}{3} \frac{\rho^2_\scrisub}{P Q_\scrisub},\\
	W^2 & \defeq   \omega^{-4}\,P\sin^2\theta \bigg[\B^2 P \sin^2\theta +\frac{\Lambda}{3}
        \bigg(\C^2 -\frac{\alpha^2 a^3}{(a^2+l^2)^2} \bigg)
            \Big[\, (a+l)^2 +(l+a\cos\theta)^2\Big] R\,\bigg],
	\end{align}
depending on $\theta$ only.	The advantage of the metric form \eqref{conformal-metric-final-alternative-1} is that one can directly use the explicit formulas in Lemma 4.1 of \cite{MPS} to compute the Cotton-York tensor of $(\,\scri,h)$, which basically coincides with the magnetic part of the rescaled Weyl tensor at $\scri$.\\

Next we also need to evaluate the {\em causal character of the conformal infinity}~$\scri$. To this end we consider the  \emph{normal} to the hypersurfaces ${\Omega=\hbox{const.}}$, ${\ct{N}{_{\alpha}}\defeq\cd{_{\alpha}}\Omega}$ . Its vector form is just defined as ${N}^{\alpha}\defeq g^{\alpha\beta}\ct{N}{_{\beta}}$. Since the differential of ${\,\Omega\,}$ given by (\ref{newOmega-q}) is ${\,\dd\Omega = \dd q + \B \sin \theta\, \dd \theta}$, and the conformal metric is (\ref{conformal-metricJP2023}), we get
	\begin{equation}
	{\bf N}=\rho^{-2}\,( Q \,\partial_{q} + \B\, P  \sin \theta\,\partial_{\theta}) \,.
	\label{eq:normal-scri}
	\end{equation}
	Its norm is ${{g}_{\alpha\beta}\,{N}^{\alpha}{N}^{\beta}=\rho^{-2}\,( Q + \B^2P\sin^2\theta)}$, and applying the identity (\ref{Q-on-scri-short}) valid on $ \scri $ it immediately follows that
	\begin{equation}
	\big(\,{g}_{\alpha\beta}\,{N}^{\alpha}{N}^{\beta}\big)_\scrisub = -\frac{\Lambda}{3}\  . \label{norm-ofn--on-scri}
	\end{equation}
	Thus (as is well known) the causal character of  $\scri$ is fully determined by the sign of the cosmological constant $\Lambda$: it is \emph{null} for ${\Lambda=0}$, \emph{spacelike} for ${\Lambda>0}$, and \emph{timelike} for ${\Lambda<0}$. Since the case of interest in the upcoming analysis is ${\Lambda>0}$, the \emph{unit timelike normal}
		\begin{equation}	\label{eq:unit-normal-scri}
			\mathbf{n}\defeq\frac{1}{\sqrt{-\ct{N}{^{\mu}}\ct{N}{_\mu}}}\,\mathbf{N}
		\end{equation}
to the hypersurfaces $ \Omega=$ const. in a neighbourhood of $ \scri $ will be used, which is future-pointing.

\subsection{Principal null directions and the Weyl and Ricci scalars}\label{ssec:PNDs}
	Applying the transformation \eqref{q=1/r} and the rescalings \eqref{rescaled-funtions} on eqs. (85) and (86) in \cite{Podolsky-Vratny:2023}, we obtain the \emph{preferred null tetrad} for the new physical metric \eqref{newmetricJP2023}, namely
	\begin{eqnarray}
	\mathbf{\hat{k}} \rovno \frac{1}{\sqrt{2}}\, \frac{\Omega}{\rho} \bigg[ \frac{1}{\sqrt{Q}}
	 \,\big(S\, \partial_t + a\,q^2 \, \partial_\varphi \big) - \sqrt{Q} \, \partial_q \bigg] \,, \nonumber \\
	\mathbf{\hat{l}} \rovno \frac{1}{\sqrt{2}}\, \frac{\Omega}{\rho} \bigg[ \frac{1}{\sqrt{Q}}
	 \,\big(S\, \partial_t + a\,q^2 \, \partial_\varphi \big) + \sqrt{Q} \, \partial_q \bigg] \,,  \label{nullframe}\\
	\mathbf{\hat{m}} \rovno \frac{1}{\sqrt{2}}\, \frac{\Omega}{\rho} \bigg[
	 \frac{1}{\sqrt{P} \sin \theta} \,\big( R\, \partial_t + \partial_\varphi \big)
	 + \mathrm{i} \, \sqrt{P} \, \partial_\theta \bigg] \,, \nonumber
	\end{eqnarray}
  where $R(\theta)$ and $S(q)$ are defined by \eqref{def-R} and \eqref{def-S}, respectively.
	In this tetrad, the only nontrivial Newman-Penrose scalar representing the \emph{Weyl curvature tensor} is
	\begin{align}
	\Psi_2 &= \frac{\Omega^3}{\big[1+\mathrm{i}\,(l+a \cos \theta)\,q\, \big]^3} \bigg[ -(m+\mathrm{i}\,l)\Big(1-\mathrm{i}\,\alpha\, a\,\frac{a^2-l^2}{a^2+l^2}\Big)
	- \mathrm{i}\, \frac{\Lambda}{3} \,l \,(a^2-l^2) \nonumber\\
	&\hspace{35mm}+\frac{(e^2+g^2)}{1-\mathrm{i}\,(l+a \cos \theta)\,q}\,
	\Big(q+\frac{\alpha\,a}{a^2 + l^2} \big[a\, \cos \theta +\mathrm{i}\, l\, (l+a \cos \theta)\,q\, \big]\Big)
	 \bigg]\,.\label{eq:psi2}
	\end{align}
	This explicitly confirms that the spacetime is of algebraic type~D, and \emph{both vectors $\mathbf{\hat{k}}$ and $\mathbf{\hat{l}}$ are the principal null directions} (PNDs). Also, with respect to \eqref{nullframe} the electromagnetic field, whose coordinate form is explicitly given  in \cref{Fab}, has the Newman-Penrose scalars ${\Phi_0 = 0 = \Phi_2}$ and
	\begin{equation}
	\Phi_1 = \frac{\frac{1}{2}(e+\mathrm{i}\,g) \,\Omega^2}{\big[1+\mathrm{i}\,(l+a \cos \theta)\,q \big]^2} \,.
	   \label{Phi1}
	\end{equation}
	Hence, the electromagnetic field is non-null, and double-aligned with both PNDs. It also follows that the Ricci tensor is ${\Phi_{11} = 2\,\Phi_1 \bar\Phi_1}$. Notice that this expression {\em evaluated on~$\scri$ depends on the acceleration} of the black hole because the condition \eqref{scri-AB}, that is ${q = \A + \B \cos \theta}$, involves $\alpha$.

	The null tetrad \eqref{nullframe} is defined only when ${Q>0}$. In order to investigate the radiative properties near (${\Lambda>0}$) conformal infinity $\scri$, which is located in the region ``above'' the cosmo-acceleration horizon ${\HH_c}$ \cite{Podolsky-Vratny:2023}, we have to consider the non-stationary region with ${Q<0}$ with the following modified version of the null tetrad
	\begin{eqnarray}
	\mathbf{\hat{k}} \rovno \frac{1}{\sqrt{2}}\, \frac{\Omega}{\rho} \bigg[ \frac{1}{\sqrt{-Q}}
	 \,\big( S\, \partial_t + a\,q^2 \, \partial_\varphi \big) + \sqrt{-Q} \, \partial_q \bigg] \,, \nonumber \\
	\mathbf{\hat{l}} \rovno \frac{1}{\sqrt{2}}\, \frac{\Omega}{\rho} \bigg[ \frac{-1}{\sqrt{-Q}}
	 \,\big( S\, \partial_t + a\,q^2 \, \partial_\varphi \big) + \sqrt{-Q} \, \partial_q \bigg] \,,  \label{nullframe-Q<0}\\
	\mathbf{\hat{m}} \rovno \frac{1}{\sqrt{2}}\, \frac{\Omega}{\rho} \bigg[
	 \frac{1}{\sqrt{P} \sin \theta} \,\big( R\, \partial_t + \partial_\varphi \big)
	 + \mathrm{i} \, \sqrt{P} \, \partial_\theta \bigg] \,. \nonumber
	\end{eqnarray}

	Interestingly, it is convenient to define the vector fields
	\begin{eqnarray}
	\textbf{T} \defeq  S\, \partial_t + a\,q^2 \, \partial_\varphi \,, \qquad
	\textbf{R} \defeq
	 \partial_\varphi + R\, \partial_t \,,  \label{TRdef}
	\end{eqnarray}
	for which,  using the identity \eqref{S-aq^2R}, it can be shown  that
	\begin{eqnarray}
	\textbf{T}\cdot\textbf{T} = -Q\,\frac{\rho^2}{\Omega^2} \,,\qquad
	\textbf{R}\cdot\textbf{R} = P \sin^2\theta\,\frac{\rho^2}{\Omega^2} \,,\qquad
	\textbf{T}\cdot\textbf{R} = 0 \,.  \label{TTTTR}
	\end{eqnarray}
	Hence,  the vector field $\textbf{T}$ is \emph{timelike} in the regions ${Q>0}$, \emph{spacelike} in the regions ${Q<0}$, and it is \emph{null} on the horizons where ${Q=0}$ (these are the Killing horizons). On the other hand, the vector field $\textbf{R}$ is \emph{everywhere spacelike}, except at ${\theta=0}$ and ${\theta=\pi}$ where its norm vanishes, defining thus geometrically the axes 
of axial symmetry. Moreover, these vector fields $\textbf{T}$  and $\textbf{R}$ are \emph{mutually orthogonal}. For the static case ${a=0=l}$ we get simply ${\textbf{T} = \partial_t}$ and ${\textbf{R} = \partial_\varphi}$, i.e. the usual generators of time translation isometry and the axial symmetry.\\
	
	Using the relations \eqref{TTTTR} it is now easy to check that both the tetrads \eqref{nullframe} and \eqref{nullframe-Q<0} satisfy the normalization conditions ${\mathbf{\hat{k}}\cdot\mathbf{\hat{l}}=-1}$ and ${\mathbf{\hat{m}}\cdot\mathbf{\bar{\hat{m}}}=1}$ (with all other scalar products vanishing). The null tetrad \eqref{nullframe-Q<0} adapted to both  double-repeated PNDs $\mathbf{\hat{k}}$ and $\mathbf{\hat{l}}$ is directly related to a
\emph{special orthonormal tetrad} ${(\mathbf{\hat{t}}, \mathbf{\hat{q}}, \mathbf{\hat{r}}, \mathbf{\hat{s}})}$
 via the usual algebraic relations
	\begin{equation}
	\mathbf{\hat{k}} \defeq  \tfrac{1}{\sqrt2} (\mathbf{\hat{t}} + \mathbf{\hat{q}})\,, \qquad
	\mathbf{\hat{l}} \defeq  \tfrac{1}{\sqrt2} (\mathbf{\hat{t}} - \mathbf{\hat{q}})\,, \qquad
	\mathbf{\hat{m}} \defeq  \tfrac{1}{\sqrt2} (\mathbf{\hat{s}} + \mathrm{i}\,
                            \mathbf{\hat{r}})\,,  \label{special-orthonormal-def}
	\end{equation}
	that is explicitly
	\begin{equation}
	 \mathbf{\hat{t}} = \frac{\Omega}{\rho} \,\sqrt{-Q} \,\, \partial_{q} \,, \qquad
	 \mathbf{\hat{q}} = \frac{\Omega}{\rho}\,\frac{1}{\sqrt{-Q}}\,\, \textbf{T} \,,  \qquad
	 \mathbf{\hat{r}} = \frac{\Omega}{\rho}\,\sqrt{P} \,\, \partial_\theta \,,  \qquad
	 \mathbf{\hat{s}} = \frac{\Omega}{\rho}\,\frac{1}{\sqrt{P}\,\sin\theta}\,\, \textbf{R} \,.
    \label{special-orthonormal}
	\end{equation}
For the metric \eqref{newmetricJP2023} these vectors are normalized  as ${\mathbf{\hat{t}}\cdot\mathbf{\hat{t}}=-1}$ and ${\mathbf{\hat{q}}\cdot\mathbf{\hat{q}}=\mathbf{\hat{r}}\cdot\mathbf{\hat{r}}=\mathbf{\hat{s}}\cdot\mathbf{\hat{s}}=1}$
(with all other scalar products vanishing).

	It can be concluded that the timelike unit vector $\mathbf{\hat{t}}$ \emph{is not collinear with the unit vector $\mathbf{n}$ \eqref{eq:unit-normal-scri}, 	 \eqref{eq:normal-scri} normal to~$\scri$, unless ${\B=0}$}, i.e. for the vanishing acceleration $\alpha$.\footnote{In order to evaluate $ \mathbf{\hat{t}} $ at $ \scri $, rescale it first with $ \Omega^{-1} $, obtaining $\rho^{-1} \,\sqrt{-Q} \,\partial_{q}$.} In fact, ${{\bf n}}$ \emph{is not coplanar} with the two PNDs $\mathbf{\hat{k}}$ and $\mathbf{\hat{l}}$, unless ${\B=0}$. This already implies that \emph{such spacetimes will have gravitational radiation arriving at $\scri$ only if the black holes are accelerating}, as follows from the discussion in \cite{Fernandez-Senovilla:2020,Fernandez-Senovilla:2022b} (see Remark IV.4 in \cite{Fernandez-Senovilla:2022b}). A detailed quantitative study of the radiative properties of these spacetimes is presented in \cref{sec:radiation}.

\subsection{Strongly oriented null tetrads adapted to the conformal boundary}\label{ssec:strong-orientation}
Let us now consider a geometrically privileged class of null tetrads adapted to a ${\Lambda>0}$ scri $\scri$ which has a spatial character. Concretely, one requires two properties:
\begin{enumerate}
\item The couple of null vectors  $\ctst{\textbf{k}}{}$ and $\ctst{\textbf{l} }{}$ span a timelike plane containing the normal $\mathbf{n}$ to $ \scri $.
\item At least one of them  is aligned with a repeated principal null direction of the re-scaled Weyl tensor $ \ct{d}{_{\alpha\beta\gamma}^\delta} $.
\end{enumerate}
Such tetrads are called \emph{strongly oriented} following the nomenclature used in \cite{Fernandez-Senovilla:2022b}. Of course, the second requirement is only possible if the spacetime is algebraically special at $ \scri $. In the particular case when the rescaled Weyl tensor $ \ct{d}{_{\alpha\beta\gamma}^\delta} $ has \emph{two different} repeated null directions (algebraic type~D) at $ \scri $, as on the spacetimes under consideration here, there is a general construction (see \cref{app:general-type-D} for full details) that can be conveniently used, as follows.

		For the unphysical metric \eqref{conformal-metricJP2023}, first rescale by $ \Omega^{-1} $ the elements of the natural  tetrad \eqref{nullframe-Q<0} and invert the orientation of $\textbf{k}$ and $\textbf{l}$ so that they become future-pointing,
	\begin{align}
		\textbf{k} &=- \frac{1}{\sqrt{2}}\, \frac{1}{\rho} \bigg[ \frac{1}{\sqrt{-Q}}
					 \,\big( S\, \partial_t + a\,q^2 \, \partial_\varphi \big) + \sqrt{-Q} \, \partial_q \bigg] \,, \nonumber \\
		\textbf{l} &=- \frac{1}{\sqrt{2}}\, \frac{1}{\rho} \bigg[ \frac{-1}{\sqrt{-Q}}
					 \,\big( S\, \partial_t + a\,q^2 \, \partial_\varphi \big) + \sqrt{-Q} \, \partial_q \bigg] \,,  \label{eq:PPN2confnullframe-Q<0}\\
		\textbf{m} &= \frac{1}{\sqrt{2}}\, \frac{1}{\rho} \bigg[
					 \frac{1}{\sqrt{P} \sin \theta}  \,\big( R\, \partial_t + \partial_\varphi \big)
					 + \mathrm{i} \, \sqrt{P} \, \partial_\theta \bigg] \,. \nonumber
	\end{align}
Next, the prescriptions \eqref{eq:boost-paramenter-b}, \eqref{eq:rotation-parameter-c} of \cref{app:general-type-D} can be applied,
giving the functions
\begin{align}
b^2&=\sqrt{1+\B^2\,\frac{P}{Q}\,\sin^2\theta}\,,\\
c  &=-\mathrm{i}\,\frac{\alpha\,a^2\,\sqrt{P}\sin\theta}
    {\sqrt{-\alpha^2 a^4\, P\sin^2\theta-(a^2+l^2)^2\, Q}}\,.
\end{align}
At $ \scri $, using the explicit form  \eqref{AB} of $ \B $ in identity \eqref{Q-on-scri-short},
the expression for $ c $ reduces to
		\begin{equation}\label{eq:param-c}
			c \eqs -\mathrm{i}\,\sqrt{\frac{3}{\Lambda}}\,\B\,\frac{\sqrt{P}\sin\theta} {\rho_\scrisub}\,,
\end{equation}
where $\rho_\scrisub (\theta)$ is given by \eqref{rho-on-scri}.  Then, the related strongly oriented tetrad is obtained by\footnote{Observe that there is no freedom to do a null rotation keeping $ \mathbf{k} $ fixed because this is fixed by the coplanarity condition.}
			\begin{align}
			\ctst{\textbf{k}}{}&=b^2\,\ct{\textbf{k}}{}\,,\nonumber\\
		 	 \ctst{\textbf{l}}{}&=\frac{1}{b^2}\,\ct{\textbf{l}}{}+c\,\ct{\textbf{m}}{}+\bar{c}\,\ct{\bar{\textbf{m}}}{}+c\bar{c}\,b^2\,\ct{\textbf{k}}{}\,, \label{eq:strong-tetrad}\\
		 	\ctst{\textbf{m}}{}&=\ct{\textbf{m}}{}+\bar{c}\,b^2\,\ct{\textbf{k}}{}\,.\nonumber
			\end{align}
Explicitly, in coordinates (where all the functions are evaluated at $ \scri $),
			\begin{align}
			 \ctst{\textbf{k}}{}&=\sqrt{\frac{\Lambda}{6}}\brkt{\frac{S}{Q}\,\partial_{t}-\partial_{q}
+\frac{\alpha^2a^3\prn{l+a\cos\theta}^2}{\prn{a^2+l^2}^2Q}\,\partial_{\phi}},\nonumber\\
			 \ctst{\textbf{l}}{}&=\sqrt{\frac{\Lambda}{6}}\Bbrkt{-\frac{S}{Q}\,\partial_{t}
-\frac{3}{\Lambda}\frac{\alpha^2a^4\,P\sin^2\theta-\prn{a^2+l^2}^2 Q}{\rho^2(a^2+l^2)^2}\,\partial_{q}
-\frac{\alpha a^2\prn{l+a\cos\theta}}{\prn{a^2+l^2}Q}\,\partial_{\phi}\nonumber\\
&\qquad\qquad+\frac{3}{\Lambda}\frac{2\alpha a^2 P\sin\theta}{\prn{a^2+l^2}^2\rho^2}\,\partial_{\theta}} ,
\label{stronly-oriented-tetrad}\\
		 	 \ctst{\textbf{m}}{}&=\frac{\sqrt{P}}{\sqrt{2}\prn{a^2+l^2}\rho}
\Bbrkt{\frac{\prn{a^2+l^2}QR+\mathrm{i}\,\alpha a^2 PS \sin^2\theta}{QP\sin\theta}\,\partial_{t}
-\mathrm{i}\,\alpha a^2 \sin\theta\,\partial_{q}\nonumber\\
&\qquad\qquad+\frac{\prn{a^2+l^2}^3Q+\mathrm{i}\,\alpha^3 a^5\prn{l+a\cos\theta}^2 P\sin^2\theta }{\prn{a^2+l^2}^2 QP\sin\theta}\partial_{\phi}+\mathrm{i}\,(a^2+l^2)\,\partial_{\theta}},\nonumber
			\end{align}
 with $R(\theta)$ and $S(\theta)$ given as
			\begin{align}
				R&= a\sin^2\theta+2l\prn{1-\cos\theta}\,,\\
				S&\eqs  1+ \alpha^2 a^2\frac{( a+l)^2}{( a^2+l^2)^2}\,\big( l + a \cos\theta \big)^2.
			\end{align}

		Observe that,  due to \eqref{eq:param-c},  a  non-vanishing $ c $ at  $\scri$ requires ${\alpha\neq 0}$. In general  asymptotic type~D scenarios, $ {c\neq 0} $ thus implies the presence of gravitational radiation. This is the case for the spacetimes considered here, and it will be shown explicitly in the following section.

\section{Gravitational radiation generated by accelerating black holes}\label{sec:radiation}
	 Asymptotically, the Weyl tensor $ \ct{C}{_{\alpha\beta\gamma}^\delta} $ of the unphysical metric \eqref{conformal-metricJP2023} vanishes --- in agreement with a general result, see for example \cite{Geroch:1977,Kroon:2016, Fernandez-Senovilla:2022b}. Hence, to study the content of gravitational radiation in the corresponding conformal spacetime~$M$, a rescaled version of this tensor,
	 	\begin{equation}
	 		\ct{d}{_{\alpha\beta\gamma}^\delta}\defeq\frac{1}{\Omega}\,\ct{C}{_{\alpha\beta\gamma}^\delta}\ ,
	 	\end{equation}
	 has to be used.\footnote{It has to be pointed out that this tensor is one of the key variables entering in the so called Conformal Einstein Field Equations --- see, e.g., \cite{Friedrich:1981b,Friedrich:1986a,Paetz:2015}.} For the metric \eqref{conformal-metricJP2023}, in the natural tetrad \eqref{eq:PPN2confnullframe-Q<0} the only non-vanishing scalar of the rescaled Weyl tensor  reads
		\begin{align}
			\phi_2 &= \frac{1}{\big[1+\mathrm{i}\,(l+a \cos \theta)\,q\, \big]^3} \bigg[ -(m+\mathrm{i}\,l)\Big(1-\mathrm{i}\,\alpha\, a\,\frac{a^2-l^2}{a^2+l^2}\Big)
			- \mathrm{i}\, \frac{\Lambda}{3} \,l \,(a^2-l^2) \nonumber\\
			&\hspace{35mm}+\frac{(e^2+g^2)}{1-\mathrm{i}\,(l+a \cos \theta)\,q}\,
			\Big(q+\frac{\alpha\,a}{a^2 + l^2} \big[a\, \cos \theta +\mathrm{i}\, l\, (l+a \cos \theta)\,q\, \big]\Big)
 \bigg]\,.\label{eq:phi2}
		\end{align}
	Notice that it is $\Psi_2$ of \cref{eq:psi2} multiplied by $ \Omega^{-3} $. This, together with the function $ c $ of \cref{eq:param-c}, can be used to compute the corresponding scalars $ \ctst{\phi}{_{4}} $, $ \ctst{\phi}{_{3}} $, and ${\ctst{\phi}{_{2}} =  \ct{\phi}{_{2}}}$ associated with the strongly oriented tetrad \eqref{eq:strong-tetrad}, \eqref{eq:PPN2confnullframe-Q<0} according to expressions \eqref{eq:sphi3} --- which are standard known formulae (see the appendices in \cite{Stewart:1991}, for instance). Then, one can use the general formulae of \cite{Fernandez-Senovilla:2022b}, or those of \cref{app:general-type-D}, to compute the electric $\ct{D}{_{ab}}$ and magnetic $ \ct{C}{_{ab}} $ parts on $\scri$ (recall the notation summarised in \cref{sec:intro})
	of the rescaled Weyl tensor in terms of $ \phi_{2} $. The \emph{matrix commutator} of these two tensors\footnote{Following the conventions of previous works \cite{Fernandez-Senovilla:2020,Fernandez-Senovilla:2022b}, the notation $ \cts{\P}{^a} $ is adopted. Observe that \eqref{eq:def-spoynting} is a real quantity; the bar should not be confused with the one used to indicate complex conjugation of complex functions, such as the rescaled Weyl scalars $ \ct{\phi}{_{i}} $, the null tetrad vector field $ \ct{m}{^\alpha} $, or the function $ c $ defined in \cref{app:general-type-D} that appears in \cref{sec:radiation-null-directions}.},
		\begin{equation}\label{eq:def-spoynting}
			\cts{\P}{^a} \defeq  2\,\epsilon^{rsa}\,\ct{C}{_{re}}\,\ct{D}{^e_{s}}\,,
		\end{equation}
  defines the \emph{asymptotic super-Poynting vector field} $ \cts{\P}{^a} $ on $\scri$.  This is the key object, as it \emph{vanishes if and only if there is no gravitational radiation at}~$ \scri $ according to the \emph{gravitational radiation condition} introduced in \cite{Fernandez-Senovilla:2022b}.
  Actually, $\cts{\P}{^a}$ is the tangential part to the spacelike $\scri$ of the \emph{canonical asymptotic supermomentum} $ \ct{\P}{^{\alpha}} $,
\begin{equation}
	   {\cal P}^{\alpha} = -\W\,n^\alpha + \cts{\P}{^a}\ct{e}{^{\alpha}_{a}} \,,
\end{equation}
   where $\W$ is the \emph{canonical asymptotic superenergy density} \cite{Fernandez-Senovilla:2020}, defined by
		\begin{equation}
			\W \defeq \ct{D}{^{rs}}\ct{D}{_{rs}}+\ct{C}{^{rs}}\ct{C}{_{rs}}\geq 0\,.
		\end{equation}
		The asymptotic supermomentum is intrinsically defined \cite{Fernandez-Senovilla:2020} from the rescaled version of the \emph{Bel-Robinson tensor} \cite{Bel:1958} (see \cite{Maartens:1998,Senovilla:2000,Alfonso:2008} and references therein for further details on these kind of tensors)
		\begin{equation}
		  \ct{\D}{_{\alpha\beta\gamma\delta}}\defeq \ct{d}{_{\alpha\mu\gamma}^\nu}\,\ct{d}{_{\delta\nu\beta}^\mu}
     +  \ctr{^*}{d}{_{\alpha\mu\gamma}^\nu}\,\ctr{^*}{d}{_{\delta\nu\beta}^\mu}\ ,
		\end{equation}
as
 $${{\cal P}^{\alpha} \defeq - \ct{\D}{^\alpha_{\beta\gamma\delta}}\,n^\beta\,n^\gamma\,n^\delta} .$$

In our case $\W$ is explicitly given by
			\begin{equation}
\W=\bigg[1-\frac{6 \alpha^2a^4 (a^2+l^2)^2 PQ\sin^2\theta}
{\big[\alpha^2a^4 P\sin^2\theta+(a^2+l^2)^2 Q\big]^2}\bigg]
\,6\phi_{2}\bar{\phi}_{2}\,,
			\end{equation}
while the spacetime version of the asymptotic super-Poynting four-vector field at $\scri$ is
		\begin{equation}\label{super-Poynting-4D}
		\cts{\P}{^{\alpha}} =  \cts{\P}{^a} e^\alpha{}_a \eqs {\cal S}(\theta)\, \big( \delta^\alpha_\theta - \B \sin\theta\,\delta^\alpha_q \big)    \,,
		\end{equation}
where
		\begin{equation}\label{def-caligraphic-S}
		  {\cal S}(\theta) \defeq
\frac{54}{\Lambda}\,\alpha\,a^2 \frac{PQ\big[\alpha^2a^4 P\sin^2\theta-(a^2+l^2)^2 Q \big]}
{\rho^3\big[-\alpha^2a^4P\sin^2\theta-(a^2+l^2)^2Q\big]^{3/2}}\, \phi_{2}\bar{\phi}_{2}\,\sin\theta \,.
		\end{equation}

It is \emph{tangential to} $ \scri $, and its explicit form for the conformal metric \eqref{conformal-metricJP2023} can be computed by the pull-back corresponding to the map ${\scri \to \partial M \subset M}$. Denoting the coordinates on~$\scri$ as ${y^a \equiv (\cts{t}{}, \cts{\varphi}{}, \cts{\theta}{})}$ and the coordinates on~$M$ as ${x^\alpha \equiv (t, \varphi, \theta, q)}$, with the natural choice ${t=\bar{t}, \varphi=\bar{\varphi}, \theta=\bar{\theta}}$ and using the relation ${q = \A + \B \cos \cts{\theta}{}}$ following from  \eqref{scri-AB}, the frame ${\ct{e}{^{\alpha}_{a}} \equiv \frac{\partial x^\alpha}{\partial y^a}}$ reads
            \begin{equation}
\ct{e}{^{\alpha}_{\bar{t}}} = \delta^\alpha_t\,,\qquad
\ct{e}{^{\alpha}_{\bar{\varphi}}} = \delta^\alpha_\varphi\,,\qquad
\ct{e}{^{\alpha}_{\bar{\theta}}}  = \delta^\alpha_\theta - \B \sin\bar{\theta}\,\delta^\alpha_q\,.\qquad
			\end{equation}
Using \eqref{super-Poynting-4D}, and rewriting $\bar{\theta}$ on $\scri$ simply as $\theta$, we thus obtain
		\begin{equation}\label{super-Poynting-3D}
\cts{\P}{^a}={\cal S}(\theta) \,\delta^a_{\theta}\,.
		\end{equation}

\begin{figure}[t!]
\centering
\includegraphics[scale=1]{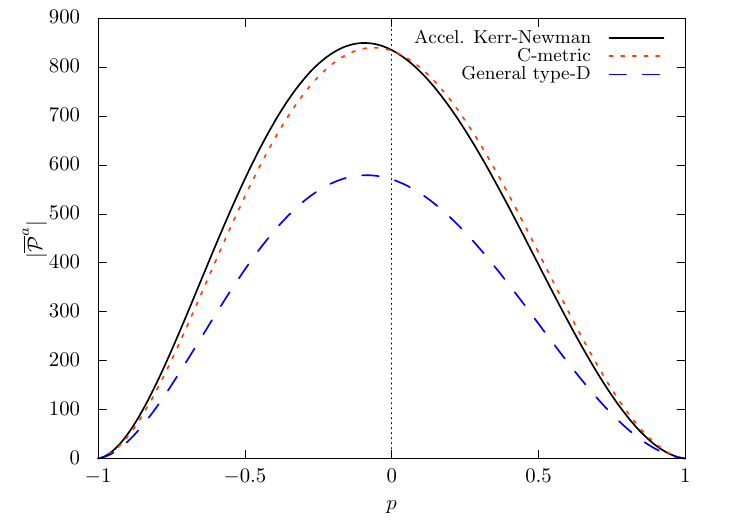}
\caption{The norm $|\cts{\P}{}|:=\sqrt{h_{ab} \cts{\P}{^a} \cts{\P}{^b}}$
of the asymptotic super-Poynting vector is plotted here as a function of $ {p\defeq\cos\theta} $. The parameters have been set such that:  ${e=g=1/8,a=1/6,l=1/20}$ for the general type-D case; $ {e=g=1/8,a=1/6,l=0} $ for the accelerating Kerr-Newman metric;  $ {e=g=0,a=0,l=0}  $ for the C-metric ($ {\Lambda=1/100} $, $ {m=\alpha=1/4} $ for all of them). Observe that the three cases feature the same type of asymmetry with respect to $ {p=0} $. This is caused by conical singularities that cannot be regularised simultaneously at both poles, and are the physical cause of the acceleration.}\label{fig:normspoynting}
\end{figure}

Moreover, one can use the identity \eqref{Q-on-scri-short} to put these key expressions to more compact forms
			\begin{eqnarray}
 \cts{\P}{^a}(\theta) \rovno 	
 \sqrt{ \frac{3}{\Lambda}}\,\frac{162}{\Lambda^2}\,\alpha\, a^2 \,
{
\frac{PQ_\scrisub\,(-Q_\scrisub+\B^2 P\sin^2\theta)}{(a^2+l^2)\,\rho^6_\scrisub}\,\phi_{2}\bar{\phi}_{2}\,\sin\theta\,\delta^a_{\theta}
}
\,, \label{eq:super-poynting-scri} \\
\W(\theta)\rovno \bigg[\,1-\frac{54}{\Lambda^2}\,\alpha^2a^4 \,\frac{PQ_\scrisub\sin^2\theta}{(a^2+l^2)\,\rho^4_\scrisub}\,\bigg]
\,6\phi_{2}\bar{\phi}_{2}\,, \label{eq:super-energy-scri}
			\end{eqnarray}
where the functions $\rho_\scrisub$, $Q_\scrisub$, $P$, evaluated at $\scri$,  are given by \eqref{rho-on-scri}, \eqref{Q-on-scri}, \eqref{P-on-scri}, respectively.\\

In this form, it is evident that $ \W $ \emph{is positive} (as it has to be), given that $ {Q<0} $  at $ \scri $  (which is always true in the non-stationary region between $\scri$ and the cosmo-acceleration horizon, see \cite{Podolsky-Vratny:2023})  and $ {P>0}. $ 
From \eqref{eq:super-poynting-scri} it is manifest that (for a generic $a$) 	
\begin{equation}
		\alpha=0 \quad\Longleftrightarrow\quad \text{ absence of gravitational radiation at $ \scri $}
\end{equation}
(recall the end of section~\ref{ssec:PNDs}). Thus, in the investigated family of exact spacetimes, \emph{only accelerating black holes generate gravitational radiation}.\\

 This agrees with the result mentioned earlier, namely that the repeated principal null directions given in \cref{eq:PPN2confnullframe-Q<0} are not coplanar with the normal to $ \scri $ \eqref{eq:normal-scri}. Equivalently, for a type~D rescaled Weyl tensor at $ \scri $, there is no gravitational radiation if and only if the two-dimensional space orthogonal to both repeated principal null directions is fully tangential to~$\scri$ (i.e. when ${c=0}$, see \cref{eq:rotation-parameter-c}).\\

 For black holes \emph{without the NUT parameter} (${l=0}$), the key quantities \eqref{eq:super-poynting-scri}, \eqref{eq:super-energy-scri} simplify to
			\begin{eqnarray}
	 \cts{\P}{^a}(\theta) \rovno \sqrt{ \frac{3}{\Lambda}}\,\frac{162}{\Lambda^2}\,
\alpha\,\frac{PQ_\scrisub\,(-Q_\scrisub + \alpha^2 P\sin^2\theta)}
{\rho^6_\scrisub}\,\phi_{2}\bar{\phi}_{2}\,\sin\theta\,\delta^a_{\theta}\,, \label{eq:super-poynting-scri-l=0} \\
\W(\theta)\rovno \bigg[\,1-\frac{54}{\Lambda^2}\,\alpha^2a^2 \,\frac{PQ_\scrisub\sin^2\theta}{\rho^4_\scrisub}\,\bigg]
\,6\phi_{2}\bar{\phi}_{2}\,. \label{eq:super-energy-scri-l=0}
			\end{eqnarray}
Again, for vanishing acceleration of such a black-hole source the asymptotic super-Poynting vector is zero, and there is no radiation at $\scri$ (located by ${q=0}$). The corresponding asymptotic superenergy density is ${\W=6m^2}$.\\

The norm 
of the asymptotic super-Poynting vector $\cts{\P}{^a}$ is plotted in \cref{fig:normspoynting} for values of $ p \defeq\cos\theta $ ranging from $ -1 $ to $ 1 $. It can be seen that in all the three cases there is a slight asymmetry with respect to $ {p=0} $. This lack of equatorial symmetry makes physical sense as there are conical singularities that cannot be regularised simultaneously at both poles/axes, so that if one regularises one pole, the other one shows a conical singularity (see \cite{Griffiths-Podolsky:2009,Podolsky-Vratny:2023} for the description of this regularisation procedure using the conicity parameter $ C $ in the range of the coordinate $ \varphi $).\\

\begin{figure}[b!]
\centering
\includegraphics[scale=1]{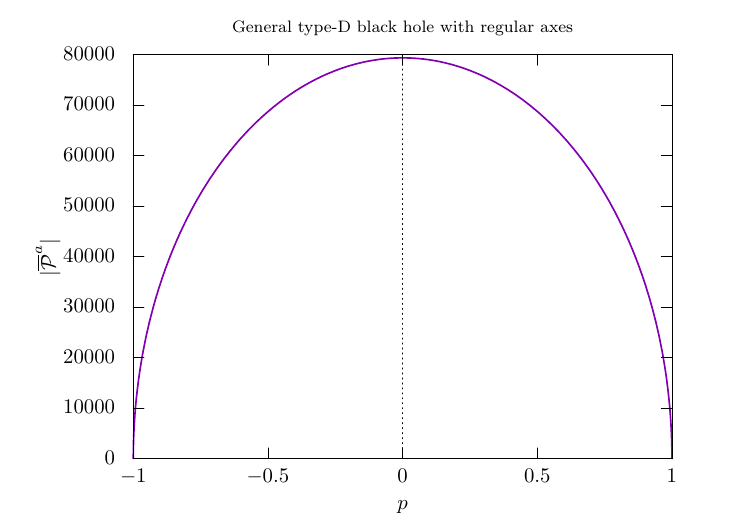}
\caption{The norm $|\cts{\P}{}|$ of the asymptotic super-Poynting vector plotted for a general black hole of type D with \emph{both axes regular}. Here the complete regularisation is achieved by ``precisely tuned'' values ${e=g=1/8,a=1/60,l=20}$, $ {m=\alpha=1/4} $ and ${\Lambda=1931475/190096}$. The curve is \emph{perfectly symmetric} with respect to $ {p=0} $.}\label{fig:normspoynting-regular}
\end{figure}

Actually, this is confirmed in \cref{fig:normspoynting-regular} where the values of the physical parameters are chosen in a very special way: they satisfy the condition given by Eq.~(190) in \cite{Podolsky-Vratny:2023} for the regularity of both axes. In this case the norm of the asymptotic super-Poynting vector is symmetric with respect to $ {p=0} $, which further reinforces the goodness of the radiation criterion employed and investigated in this work.\\

 It can be noticed that the super-Poynting vector \eqref{eq:super-poynting-scri} also vanishes for ${a=0}$. However, in view of the recent works \cite{Astorino:2024, Astorino:arXiv}, this seems to be rather an effect of degeneracy of the parameters $\alpha$ and $a$ in the general metric \eqref{newmetricJP2023}. The acceleration parameter $\alpha$ always appears to be multiplied by the spin parameter $a$, so that ${a=0}$ effectively removes $\alpha$ from the metric. Therefore, accelerating purely NUT black holes are \emph{not} described by the metric \eqref{newmetricJP2023}, although such black holes exist in the whole family of type~D spacetimes. Clarification of this subtle point will be presented elsewhere.

	\subsection{Further characterisation of radiation using null directions}\label{sec:radiation-null-directions}
		It is also possible to  further characterise the gravitational radiation arriving at $ \scri $ by associating its components to a null tetrad \cite{KrtousPodolsky:2004}. In the language of \cite{Fernandez-Senovilla:2022a,Fernandez-Senovilla:2022b}, this can be done in connection with the radiation condition by using the so called \emph{radiant} superenergy quantities, i.e, lightlike projections of the rescaled Bel-Robinson tensor. These objects  follow from the definition of the radiant supermomentum for a \emph{particular future lightlike vector field} $ \mathbf{l} $,
			\begin{equation}\label{eq:def-radiant-smomentum}
	 \ctell{\Q}{^\alpha}\defeq-\ct{\D}{^{\alpha}_{\mu\nu\rho}}\,\ct{l}{^\mu}\ct{l}{^\nu}\ct{l}{^\rho}\,\,,
			\end{equation}
whose components, denoted by
\begin{equation}
\ctell{\W}{}\defeq -\ct{l}{^\mu}\ctell{\Q}{_{\mu}}  \qquad \hbox{and} \qquad
\ctell{\Z}{}\defeq -\ct{k}{^\mu}\ctell{\Q}{_{\mu}} \,,
\end{equation}
correspond to $ \phi_{4}\bar{\phi}_{4} $ and $ \phi_{3}\bar{\phi}_{3} $ in a null basis using  ${(\mathbf{k},\mathbf{l})}$ as the null vector fields --- see section~2 of \cite{Fernandez-Senovilla:2022b} for general definitions and useful formulae. In addition, one has
\begin{equation}
		 \ctell{\Q}{^\alpha}=0 \quad\iff\quad \ctell{\W}{}=0=\ctell{\Z}{}  \quad\iff\quad \phi_{4}=0=\phi_{3}\ .
\end{equation}
		
		 In this context, there are \emph{two convenient null tetrads} that we can use. One of them ${(\mathbf{\ctst{{k}}{}},\mathbf{\ctst{{l}}{}},\mathbf{\ctst{{m}}{}})} $ is the strongly oriented tetrad \eqref{stronly-oriented-tetrad}; the other one $ {(\mathbf{\ctst{\tilde{k}}{}},\mathbf{\ctst{\tilde{l}}{}},\mathbf{\ctst{\tilde{m}}{}})} $  has $ \ctst{\tilde{\mathbf{l}}}{} $ as the other (so far not used) repeated principal null direction --- observe that the corresponding $ \ctst{\tilde{\mathbf{k}}}{} $ is then rotated with respect to \cref{eq:PPN2confnullframe-Q<0} and no longer aligned with a PND. Using the results from \cref{app:general-type-D}, this new strongly oriented tetrad $ {(\ctst{\tilde{\mathbf{k}}}{},\ctst{\tilde{\mathbf{l}}}{},\ctst{\tilde{\mathbf{m}}}{})} $ can be constructed by computing the corresponding parameters of  \cref{eq:rotation-parameter-c,eq:boost-parameter-other-strong-orientation},

\begin{align}
\tilde{b}^2&=\sqrt{1+\B^2\,\frac{P}{Q}\,\sin^2\theta} \ ,\\
c  &=-\mathrm{i}\,\frac{\alpha\,a^2\,\sqrt{P}\,\sin\theta} {(a^2+l^2)\,\sqrt{-Q-\B^2 P\sin^2\theta}}\,,
\end{align}
respectively. Observe that for this particular case $ {\tilde{b}=b} $, but this is not necessarily the case in general. Then, the radiant superenergy quantities can be computed for \emph{both} strong orientations:
				\begin{align}
			 		 \ctl{\Z}{}&=\ctkt{\Z}{}=\frac{108}{\Lambda}\,\B^2 \,\frac{P\sin^2\theta}{\rho^2}\,\ct{\phi}{_{2}}\bar{\ct{\phi}{_{2}}}\,,\\
			 		 \ctl{\W}{}&=\ctkt{\W}{}=\frac{1296}{\Lambda^2}\,\B^4 \,\frac{P^2\sin^4\theta}{\rho^4}\,\ct{\phi}{_{2}}\bar{\ct{\phi}{_{2}}}\, ,\\
			 		 \ctk{\W}{}&=0=\ctk{\Z}{}\,,\qquad
			 		 \ctlt{\W}{}=0=\ctlt{\Z}{}\,,
			 	\end{align}
and
                \begin{align}
\ctkt{\Q}{^\alpha}&=-\frac{54}{\Lambda}\frac{\sqrt{2}\,\alpha^2a^4P\sin^2\theta}{(a^2+l^2)^3\rho^3}\,\ct{\phi}{_{2}}\bar{\ct{\phi}{_{2}}}
\Bigg[-\frac{S}{Q}\,\delta^\alpha_{t}+\frac{3}{\Lambda}\frac{\alpha^2a^4P\sin^2\theta-(a^2+l^2)^2Q}{(a^2+l^2)^2\rho^2}\,\delta^\alpha_{q}\nonumber\\
			 		 &\hspace{50mm}-\frac{\alpha^2a^3(l+a\cos\theta)^2}{(a^2+l^2)^2Q}\delta^\alpha_{\varphi}+\frac{6}{\Lambda}\frac{\alpha\, a^2P\sin\theta}{(a^2+l^2)\rho^2}\,\delta^\alpha_{\theta}\Bigg], \\
			 		 \ctl{\Q}{^\alpha}&=-\frac{54}{\Lambda}\frac{\sqrt{2}\,\alpha^2a^4P\sin^2\theta}{(a^2+l^2)^3\rho^3}\,\ct{\phi}{_{2}}\bar{\ct{\phi}{_{2}}}
\Bigg[\frac{S}{Q}\,\delta^\alpha_{t}+\frac{3}{\Lambda}\frac{\alpha^2a^4P\sin^2\theta-(a^2+l^2)^2Q}{(a^2+l^2)^2\rho^2}\,\delta^\alpha_{q}\nonumber\\
			 		 &\hspace{50mm}+\frac{\alpha^2a^3(l+a\cos\theta)^2}{(a^2+l^2)^2Q}\,\delta^\alpha_{\varphi}+\frac{6}{\Lambda}\frac{\alpha\, a^2P\sin\theta}{(a^2+l^2)\rho^2}\,\delta^\alpha_{\theta}\Bigg].
			 	\end{align}
Again, these radiation quantities vanish when ${\alpha=0}$ (implying ${\B=0}$), that is in the absence of acceleration.\\

			The related \emph{geometrical parameter} $ \beta $, defined in \cref{eq:beta} of \cref{app:general-type-D}, gives the \emph{relative orientation} between the spatial parts of the radiant supermomenta $\ctl{\Q}{^\alpha} $ and $ \ctkt{\Q}{^\alpha} $  with respect to $ \scri $. Intuitively, this can be interpreted as measuring the angle between the spatial components of the `tidal momentum' of gravitational waves as experienced by an observer oriented along $ \ctst{\mathbf{l}}{} $, with respect to the detection of a different observer aligned with $ \ctst{\mathbf{\tilde{k}}}{} $.  In a fully general case representing the whole family of type~D black holes, one gets the expression
	\begin{align}\label{beta-parameter}
		\beta = 1 - \frac{2\prn{Q_\scrisub+\B^2P\sin^2\theta}^3}{Q_\scrisub \prn{Q_\scrisub -3\B^2P\sin^2\theta}^2}\,.
	\end{align}
Without acceleration (${\alpha=0}$, so that ${\B=0}$), this parameter reduces to
				\begin{align}\label{beta-parameter-no-acceleration}
					\beta =-1\,,
				\end{align}
while for black holes without the NUT parameter (${l=0}$, implying ${\B=\alpha}$) we get
	\begin{align}\label{beta-parameter-no-NUT}
		\beta = 1 - \frac{2\,{\big( Q_\scrisub +\alpha^2 P\sin^2\theta \big)}^3}
{ Q_\scrisub \,{\big( Q_\scrisub  - 3\alpha^2 P \sin^2\theta \big)}^2}\,.
	\end{align}

In particular, for the \emph{C-metric} without rotation (${a=0}$) and charges (${e=0=g}$), the metric functions on $ \scri $ take the form
	\begin{eqnarray}
P = 1 - 2\,\alpha m \cos\theta \,,\qquad
Q_\scrisub = -\frac{\Lambda}{3} - \alpha^2 P \sin^2\theta\,,
	\end{eqnarray}
so that
\begin{align}
c \eqs -\mathrm{i}\,\alpha\,\sqrt{\frac{3}{\Lambda}}\,\sqrt{P}\sin\theta \,.
\end{align}

\begin{figure}[h!]
			\centering
			\includegraphics[scale=1]{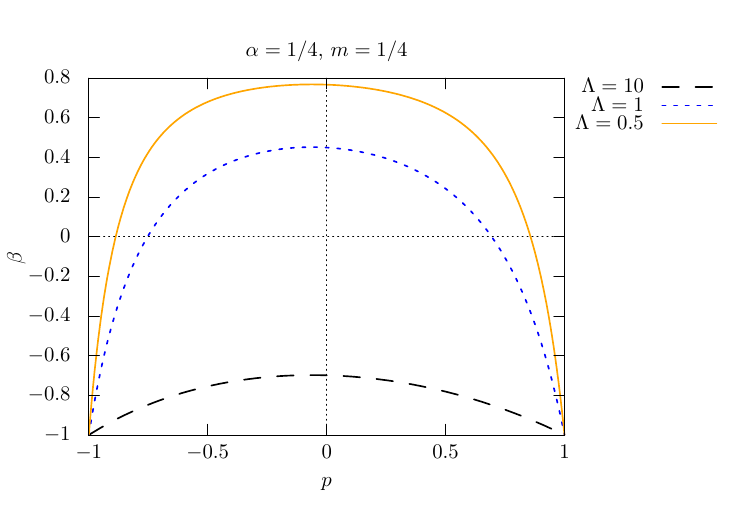}
			\caption{The parameter $ \beta $ of \cref{eq:beta} for the C-metric with fixed values of the mass $ {m=1/4} $, the acceleration $ {\alpha=1/4} $, and three distinct values of the cosmological constant $ {\Lambda>0} $. Depending on the strength of $ \Lambda $, the angular parameter $ \beta $ takes different values in $ \scri $. If $ \Lambda $ is sufficiently large, $ \beta $ is always negative.   Notice also a slight asymmetry of the curves with respect to ${p=0}$. This is the same feature observed in the norm of the asymptotic super-Poynting vector in \cref{fig:normspoynting}. }\label{fig:beta-cmetric}
			\end{figure}

In \cref{fig:beta-cmetric}, the parameter $\beta$ given by \eqref{beta-parameter-no-NUT} is plotted as a function of  $ {p \defeq \cos\theta} $ for fixed values of $\alpha$ and $m$, and three values of the positive cosmological constant~$\Lambda$. As one can see, there are \emph{three possible regimes}, described in \cref{app:general-type-D}, depending on whether $ \beta $ is positive (the projected radiant supermomenta form an acute angle), negative (the projected radiant supermomenta form an obtuse angle) or zero (the projected radiant supermomenta form exactly a $ \pi/2 $ angle).

A discussion of the gravitational radiation propagating along the PNDs $ \ctst{\mathbf{k}}{} $ and $ \ctst{\mathbf{\tilde{l}}}{} $, emitted by the accelerating black-holes, represented by the C-metric with ${\Lambda>0}$, can be found in \cite{Fernandez-Senovilla:2022b} and, in relation to the directional structure of radiation, also in \cite{KrtousPodolsky:2003tc,KrtousPodolsky:2004}.

\section{Relation to asymptotic directional structure of radiation}
\label{asymptotic-direction-structure}

Finally, we can relate these results to our previous studies concerning the structure of radiation in spacetimes with any cosmological constant~$\Lambda$, presented in
\cite{KrtousPodolsky:2004, KrtousPodolskyBicak:2003rw,KrtousPodolsky:2005rd, Podolsky-Kadlecova:2009}.

\subsection{Supermomentum vector orientation on $\scri$ with ${\Lambda>0}$}
\label{supermomentum-vector}
Recall that the \emph{asymptotic super-Poynting four-vector} $ \cts{\P}{^{\alpha}} $ is explicitly given by \eqref{super-Poynting-4D}, namely
			\begin{eqnarray}
	 \cts{\mathbf{P}}{} \eqs {\cal S}\, \big( \,{\partial}_{\theta} - \B \sin\theta\,\,{\partial}_q \big) \,, \label{eq:super-poynting-scri-repeated}
			\end{eqnarray}
where ${\cal S}(\theta)$ is given by  \eqref{def-caligraphic-S} and ${\B = \alpha\,a^2/(a^2+l^2)}$, see \eqref{AB}. This key radiation quantity can be expressed with respect to any  null tetrad. In particular, in \cite{Fernandez-Senovilla:2022b} an equivalent statement of the radiation condition was presented in terms of the rescaled-Weyl scalars in any null tetrad \emph{coplanar with the normal to} $ \scri $, namely
\begin{align}
	8\ct{\phi}{_1}\ct{\bar{\phi}}{_1}-8\ct{\phi}{_3}\ct{\bar{\phi}}{_3}-4\ct{\phi}{_4}\ct{\bar{\phi}}{_4}+4\ct{\phi}{_0}\ct{\bar{\phi}}{_0} &= 0\, ,\\
	\ct{\phi}{_3}\ct{\bar{\phi}}{_4}+\ct{\phi}{_0}\ct{\bar{\phi}}{_1}-3\ct{\phi}{_1}\ct{\overline{\phi}}{_2}-3\ct{\phi}{_2}\ct{\overline{\phi}}{_3}& = 0 \,.
\end{align}
Indeed, if a strong orientation such as \cref{eq:strong-tetrad} is chosen, then the radiation condition for asymptotically type-D spacetimes reduces to
$$ \cts{\P}{^a}=0 \quad\iff\quad \phi_{4}=0=\phi_{3} \, \, \mbox{(under strong orientation)}.$$

As a next task, we aim to elucidate the relation between the spatial direction of the asymptotic super-Poynting vector and the maximum in the directional pattern of the dominant $ \Psi_4^i $ scalar on $\scri$ derived for any type D spacetime in a suitable basis called `interpretation tetrad' in \cite{KrtousPodolsky:2004}. To do so, it is convenient to start from the \emph{algebraically privileged null tetrad} \eqref{nullframe-Q<0} for which \emph{both vectors $\mathbf{\hat{k}}$ and~$\mathbf{\hat{l}}$ are the double-repeated principal null directions} of the type~D spacetime. In order to employ a frame that is closest to the strongly oriented null tetrad adapted to the conformal boundary~$\scri$ --- as introduced in subsection~\ref{ssec:strong-orientation} --- we will consider the null frame \eqref{eq:PPN2confnullframe-Q<0} in the unphysical (conformal) spacetime \eqref{conformal-metricJP2023}. Such a frame is obtained from \eqref{nullframe-Q<0} via ${\mathbf{k}=-\mathbf{\hat{k}}/\Omega}$, ${\mathbf{l}=-\mathbf{\hat{l}}/\Omega}$, ${\mathbf{m}=\mathbf{\hat{m}}/\Omega}$. The associated \emph{special orthonormal tetrad} ${(\mathbf{t}, \mathbf{q}, \mathbf{r}, \mathbf{s})}$,
is obtained using the relations
	\begin{equation}
{\mathbf{k} = \tfrac{1}{\sqrt2} (\mathbf{t} + \mathbf{q})}\,,\qquad
{\mathbf{l} = \tfrac{1}{\sqrt2} (\mathbf{t} - \mathbf{q})}\,,\qquad
{\mathbf{m} = \tfrac{1}{\sqrt2} (\mathbf{s} + \mathrm{i}\,\mathbf{r})}\,. \label{null-to-orthonormal}
	\end{equation}
Notice that the role of the transversal spatial vectors  $\mathbf{r}$, $\mathbf{s}$ and $\mathbf{\hat{r}}$, $\mathbf{\hat{s}}$, as introduced in \eqref{special-orthonormal-def}, are swapped with respect to those employed in \cite{KrtousPodolsky:2005rd,Podolsky-Kadlecova:2009}  --- where such a  special tetrad is denoted as ${(\mathbf{t}_{\rm s}, \mathbf{q}_{\rm s}, \mathbf{r}_{\rm s}, \mathbf{s}_{\rm s})}$. \footnote{Such a swapping corresponds to the change $\mathbf{m}  \leftrightarrow - \mathrm{i}\,\mathbf{m} $.} This tetrad here reads
	\begin{equation}
	 \mathbf{t} = -\frac{1}{\rho} \,\sqrt{-Q} \,\, \partial_{q} \,, \quad
	 \mathbf{q} = -\frac{1}{\rho}\,\frac{1}{\sqrt{-Q}}\,\, \textbf{T} \,,  \quad
	 \mathbf{r} = \frac{1}{\rho}\,\sqrt{P} \,\partial_\theta \,,  \quad
	 \mathbf{s} = \frac{1}{\rho}\,\frac{1}{\sqrt{P}\,\sin\theta}\,\, \textbf{R} \,,
\label{special-orthonormal-repeated}
	\end{equation}
in which the spacelike vector field ${\textbf{T} =  S\, \partial_t + a\,q^2 \, \partial_\varphi}$ and the spacelike vector field
${\textbf{R} = \partial_\varphi + R\, \partial_t}$ are mutually orthogonal, see \eqref{special-orthonormal-def}, \eqref{special-orthonormal}, \eqref{TRdef}, \eqref{TTTTR}. The vectors \eqref{special-orthonormal-repeated} are normalized as
${\mathbf{t}\cdot\mathbf{t}=-1}$ and ${\mathbf{q}\cdot\mathbf{q}=\mathbf{r}\cdot\mathbf{r}=\mathbf{s}\cdot\mathbf{s}=1}$ (all other scalar products vanish).

Expressed in terms of this tetrad, the asymptotic super-Poynting four-vector \eqref{eq:super-poynting-scri-repeated} has the form
			\begin{eqnarray}
	 \cts{\mathbf{P}}{} = \frac{{\cal S}\,\rho_\scrisub}{\sqrt{P}}\, \Big( \,\mathbf{r} + \B\,\sqrt{\frac{P}{-Q_\scrisub}}\sin\theta\,\,\mathbf{t} \Big) \,.
\label{eq:super-poynting-scri-in-special frame}
			\end{eqnarray}

			\begin{figure}[t!]
			\centering
			\includegraphics[scale=1.8]{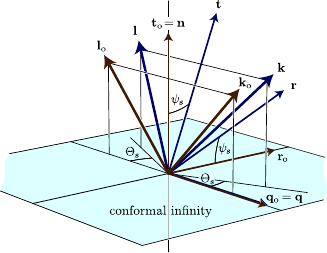}
			\caption{The scheme of various frames, in particular the orthonormal reference tetrad ${(\mathbf{t}_{\rm o}, \mathbf{q}_{\rm o}, \mathbf{r}_{\rm o}, \mathbf{s}_{\rm o})}$ and the orthonormal tetrad ${(\mathbf{t}, \mathbf{q}, \mathbf{r}, \mathbf{s})}$ associated to the PNDs $\mathbf{k}$ and $\mathbf{l}$. Their relation is given by the rapidity parameter $\psi_{\rm s}$, uniquely corresponding to the angle $\Theta_{\rm s}$. Coplanarity with the unit normal~$\mathbf{n}$ to the conformal infinity~$\scri$ can be seen and understood.}\label{fig:figure-directions-on-scri}
			\end{figure}

By comparing the timelike vectors ${\mathbf{t} \>\propto\>  \partial_{q}}$ and ${\mathbf{n} \> \propto\>   Q \,\partial_{q} + \B\, P  \sin \theta\,\partial_{\theta}}$ (given by  \eqref{special-orthonormal-repeated} and \eqref{eq:unit-normal-scri}, respectively) it is now obvious that ${\mathbf{t}}$ \emph{is not normal to}~$\scri$, unless ${\B=0}$. Because ${\mathbf{t}} \> \propto\>  \mathbf{k} + \mathbf{l}$, and as stated in \cref{sec:radiation}, this means that the plane spanned by the two (double-degenerate) principal null directions $\mathbf{k}$ and $\mathbf{l}$ \emph{is not perpendicular to} $\scri$ when the black holes are accelerating (${\alpha\not=0}$) and generate gravitational radiation.\\

To determine the \emph{spatial} direction on $\scri$ in which the gravitational radiation propagates --- that is to geometrically identify the direction of the asymptotic super-Poynting vector tangential to the three-dimensional conformal infinity~$\scri$--- it is useful to choose another orthonormal tetrad ${(\mathbf{t}_{\rm o}, \mathbf{q}_{\rm o}, \mathbf{r}_{\rm o}, \mathbf{s}_{\rm o})}$, which may be called a \emph{reference tetrad}. It is defined by the property
that the timelike (unit) vector ${\mathbf{t}_{\rm o}}$ is \emph{collinear with} the normal ${{\bf n}}$ to $\scri$. Such a reference tetrad can be obtained from the special orthonormal tetrad \eqref{special-orthonormal-repeated} adapted to both PNDs by a suitable Lorentz transformation, namely the \emph{simple boost} in the ${(\mathbf{t},\mathbf{r})}$ plane
\begin{eqnarray}
\mathbf{t}_{\rm o} \rovno
    \cosh\psi_{\rm s}\, \mathbf{t} + \sinh\psi_{\rm s}\, \mathbf{r} \,, \nonumber \\[1mm]
\mathbf{r}_{\rm o} \rovno
    \sinh\psi_{\rm s}\, \mathbf{t} + \cosh\psi_{\rm s}\, \mathbf{r} \,,  \label{reference-special-rotation}\\[1mm]
\mathbf{q}_{\rm o} \rovno \mathbf{q} \,,  \qquad
\mathbf{s}_{\rm o} = \mathbf{s} \,,  \nonumber
\end{eqnarray}
where the \emph{rapidity parameter}~$\psi_{s}$ has a special value
\begin{equation}
\tanh\psi_{\rm s} \defeq  \B\, \sqrt{\frac{P}{-Q}}\,\sin \theta\,,  \label{psi-special}
\end{equation}
which using the  identity \eqref{Q-on-scri-short} on $\scri$ implies
\begin{equation}
\cosh\psi_{\rm s} = \sqrt{\frac{3}{\Lambda}}\,\frac{\sqrt{-Q}}{\rho}\,.  \label{psi-special-equiv}
\end{equation}
Therefore,
\begin{eqnarray}
\mathbf{t}_{\rm o} \rovno \sqrt{\frac{3}{\Lambda}}\,\frac{\sqrt{-Q}}{\rho} \,
\bigg( \mathbf{t} + \B\,\sqrt{\frac{P}{-Q}}\,\sin \theta\,\, \mathbf{r} \bigg) , \nonumber \\
\mathbf{r}_{\rm o} \rovno \sqrt{\frac{3}{\Lambda}}\,\frac{\sqrt{-Q}}{\rho} \,
\bigg( \mathbf{r} + \B\,\sqrt{\frac{P}{-Q}}\,\sin \theta\,\, \mathbf{t} \bigg) ,  \label{reference-special-transf} \
\end{eqnarray}
and the reference tetrad in the coordinate frame explicitly reads
\begin{eqnarray}
\mathbf{t}_{\rm o} \rovno \sqrt{\frac{3}{\Lambda}}\,\frac{1}{\rho^2} \,\big( Q \,\partial_{q} + \B\, P  \sin \theta\,\partial_{\theta}\big) \,, \nonumber \\
\mathbf{r}_{\rm o} \rovno \sqrt{\frac{3}{\Lambda}}\,\frac{1}{\rho^2}\,\sqrt{-PQ} \,\big( \partial_{\theta} - \B  \sin \theta\,\partial_q \big)\,,  \label{reference-orthonormal}\\
\mathbf{q}_{\rm o} \rovno - \frac{1}{\rho}\,\frac{1}{\sqrt{-Q}}\, \textbf{T} \,,  \qquad
\mathbf{s}_{\rm o} = \frac{1}{\rho}\,\frac{1}{\sqrt{P}\,\sin\theta}\, \textbf{R} \,.\nonumber
\end{eqnarray}
It is now obvious from \eqref{eq:unit-normal-scri} that
\begin{eqnarray}
   \mathbf{t}_{\rm o} \rovno {\bf n} \,,  \label{t0-n}
\end{eqnarray}
as plotted in \cref{fig:figure-directions-on-scri}, and the spatial unit vectors ${(\mathbf{q}_{\rm o}, \mathbf{r}_{\rm o}, \mathbf{s}_{\rm o})}$ span the tangent space to $\scri$. Moreover, in view of the relation \eqref{reference-special-transf}, the \emph{asymptotic super-Poynting four-vector} \eqref{eq:super-poynting-scri-in-special frame} expressed in the \emph{reference tetrad} is
			\begin{eqnarray}
	 \cts{\mathbf{P}}{} = \sqrt{\frac{\Lambda}{3}} \,\frac{{\cal S}\,\rho^2_\scrisub}{\sqrt{-PQ_\scrisub}}\,\, \mathbf{r}_{\rm o} \,.
\label{eq:super-poynting-scri-in-reference frame}
			\end{eqnarray}
This is \emph{the unique spatial direction of propagation of the gravitational radiation} in the reference frame. Notice that it is geometrically privileged, because at any point on $\scri$ it is the \emph{intersection of the conformal infinity~$\scri$ with the plane spanned by $\mathbf{t}$ and $\mathbf{r}$}, which is the ``symmetry plane'' of the two PNDs $\mathbf{k}$~and~$\mathbf{l}$, see \cref{fig:figure-directions-on-scri}.

As discussed in \cref{sec:radiation}, in the absence of gravitational radiation the plane spanned by the two (double-degenerate) principal null directions $\mathbf{k}$ and $\mathbf{l}$ is perpendicular to $\scri$ (${\psi_{\rm s}=0}$). Indeed, there is the following general relation between the fundamental complex function $ c $ of \cref{eq:rotation-parameter-c} and the rapidity parameter $\psi_{\rm s}$:
	\begin{equation}\label{eq:c-psi}
		c = -\mathrm{i}\,\sinh\psi_{\rm s}\,.
	\end{equation}
Using \cref{psi-special} is easy to check that \cref{eq:c-psi} gives \cref{eq:param-c}. There is thus no radiation  when ${c=0\ \Leftrightarrow\ \psi_{\rm s}=0}$. Also, the asymptotic super-Poynting vector $\cts{\mathbf{P}}{}$ vanishes when ${{\cal S}=0\ \Leftrightarrow\ \alpha=0}$ .

\subsection{Asymptotic structure of radiation in general type D spacetimes with ${\Lambda>0}$}
\label{asymptotic-structure-general-D}

Finally, we can relate the four-dimensional geometric scheme of \cref{fig:figure-directions-on-scri} to the \emph{directional spatial structure of the specific Weyl scalar $ {\Psi }_4^{\rm i} $} at any point on~$\scri$, summarised in \cite{KrtousPodolsky:2004}. To this end, let us first observe that the rapidity parameter $\psi_{\rm s}$ of the boost is uniquely related to the rotational parameter $\Theta_{\rm s}$ in the spatial section of $\scri$ spanned by the vectors ${(\mathbf{q}_{\rm o},\mathbf{r}_{\rm o})}$ as
\begin{equation}
\tanh\psi_{\rm s} = \sin\Theta_{\rm s}
\quad\Longleftrightarrow\quad
\sinh\psi_{\rm s} = \tan\Theta_{\rm s}
\quad\Longleftrightarrow\quad
\cosh\psi_{\rm s} = \frac{1}{\cos\Theta_{\rm s}} \,.
\label{psi-special-theta-special}
\end{equation}
Clearly, ${\psi_{\rm s} = 0}$ corresponds to ${\Theta_{\rm s} = 0}$, which is in the absence of acceleration, and thus no radiation.

	\begin{figure}[b!]
			\centering

 		\includegraphics[scale=1.3]{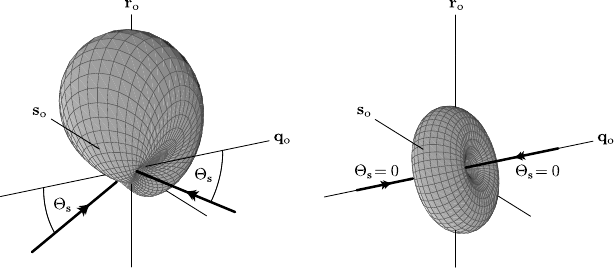}
\caption{
Directional pattern  $\mathcal{A}(\Theta,\Phi)$, for a fixed $\Theta_{\rm s}$, of $ \Psi_4^{\rm i} $ which is the dominant Weyl component at~$\scri$, as given by \cref{asyfield1+} in a generic type~D spacetime with ${\Lambda>0}$.  It applies to the class of black hole solutions studied in this
paper, in particular to the C-metric (left, ${\Theta_{\rm s}\ne0}$) and the Schwarzschild-de~Sitter solution (right, ${\Theta_{\rm s}=0}$). These plots \emph{alone} do not demonstrate the presence/absence of gravitational radiation at~$ \scri $ because $ \Psi_4^{\rm i} $ represents the asymptotic value of the gravitational field with respect to arbitrarily oriented interpretation  null tetrad, parallel propagated along various null geodesics reaching the given point at conformal infinity. However, they clearly show that $ \Psi_4^{\rm i} $ has its \emph{maximal value along the spatial direction}~$\mathbf{r}_{\rm o}$ --- which is caused by the asymmetry with respect to the plane ${(\mathbf{q}_{\rm o}, \mathbf{s}_{\rm o})}$ when ${\Theta_{\rm s} \ne 0}$, that is for accelerating black holes. According to \eqref{eq:super-poynting-scri-in-reference frame},~$\mathbf{r}_{\rm o}$ is \emph{exactly the direction of the asymptotic super-Poynting vector} $\cts{\mathbf{P}}{}$.\label{fig:figure-type-D-radiation-pattern}
 }
			\end{figure}

In fact, \emph{any unit spatial direction $\mathbf{v}$ on the} ${\Lambda>0}$ \emph{scri}~$\scri$ can by parameterized by \emph{two spherical angles $\Theta$ and $\Phi$} via the standard relation
\begin{eqnarray}
\mathbf{v} \rovno \cos\Theta\,\mathbf{q}_{\rm o} + \sin\Theta\,(\cos\Phi\,\mathbf{r}_{\rm o} + \sin\Phi\,\mathbf{s}_{\rm o}) \,. \label{def-angles-on-scri} \
\end{eqnarray}
Using such a framework and formalism, in \cite{KrtousPodolsky:2004, KrtousPodolsky:2003tc, KrtousPodolskyBicak:2003rw} --- and  in \cite{KrtousPodolsky:2005rd}  specifically in more detail for type~D spacetimes --- the  formula
\begin{equation}
|{\Psi }_4^{\rm i}| = \frac{3}{2|\eta|}\frac{|\phi_2|}{\cos^2\Theta_{\rm s}}\,\mathcal{A}(\Theta,\Phi,\Theta_{\rm s})\,\label{asyfield1+}
\end{equation}
was derived, where ${\Psi }_4^{\rm i}$ is the dominant part of the Weyl tensor expressed in a parallel-propagated ``interpretation'' null frame approaching $\scri$,
$\eta$ is an affine parameter along the concrete null geodesic ending at a given point at $\scri$, 
and
\begin{equation}
\mathcal{A}(\Theta,\Phi,\Theta_{\rm s}) \defeq (\sin\Theta+\sin\Theta_{\rm s}\cos\Phi)^2 +\sin^2\Theta_{\rm s}\cos^2\!\Theta\sin^2\!\Phi \label{amplitudseA}
\end{equation}
is the specific  function determining the directional structure of $ \Psi_4^{\rm i} $ in any type~D spacetime with the $\Lambda>0$  conformal infinity~$\scri$. Such a directional structure is absent in asymptotically flat spacetimes.

This pattern is represented on Fig.~\ref{fig:figure-type-D-radiation-pattern} (adapted from \cite{KrtousPodolsky:2003tc, KrtousPodolsky:2005rd}). The \emph{global maximum} of $\mathcal{A}$ occurs for ${\Theta=\pi/2}$, ${\Phi=0}$, that is \emph{along the spatial direction}~${\mathbf{v} = \mathbf{r}_{\rm o}}$. \emph{This is consistent with} \eqref{eq:super-poynting-scri-in-reference frame} determining the spatial direction of the asymptotic supermomentum vector $\cts{\mathbf{P}}{}$ at~$\scri$.\\

Let us also recall another result reviewed in \cite{KrtousPodolsky:2004, KrtousPodolsky:2005rd}, namely that \emph{there is no radiation arriving at $\scri$ from the spatial directions exactly opposite to principal null directions}. On Fig.~\ref{fig:figure-type-D-radiation-pattern}, the spatial projections on $\scri$ of the double-degenerate PNDs are indicated by two bold ``incoming'' double arrows ${\Theta=\Theta_{\rm s}}$, ${\Phi=0}$, and ${\Theta=\pi-\Theta_{\rm s}}$, ${\Phi=0}$. The zeros of the function \eqref{amplitudseA} occur at ${\Theta=\pi-\Theta_{\rm s}}$, ${\Phi=\pi}$, and  ${\Theta=\Theta_{\rm s}}$, ${\Phi=\pi}$, which are the two antipodal spatial directions to these PNDs.

As shown in \cite{KrtousPodolsky:2005rd}, this picture applies to any spacetime of algebraic type~D 
with a spacelike $\scri$. A specific exact solution of this type, namely the C-metric with ${\Lambda>0}$, was studied in considerable detail \cite{KrtousPodolsky:2003tc} proving that the zeros of the function $\mathcal{A}$ geometrically correspond to the \emph{two spatial directions} which are \emph{opposite} to the directions of accelerating pair of black holes, \emph{as observed from}~$\scri$. Such an analysis was subsequently generalized in \cite{Podolsky-Kadlecova:2009} to a complete family of accelerating, charged and rotating black holes with any value of the cosmological constant (and double-aligned non-null electromagnetic field), which belong to the type~D class spacetimes. In fact, the metrics and the frames employed here, and those used previously, are closely related. (For example, the tetrads  \eqref{nullframe-Q<0},  \eqref{special-orthonormal} are analogues of eqs.~(13), (28) in \cite{Podolsky-Kadlecova:2009}, the Weyl scalar \eqref{eq:psi2} corresponds to eq.~(14) therein, etc.) This last conclusion is also in agreement with  \cite{Fernandez-Senovilla:2022b,Fernandez-Senovilla:2020} where the radiant supermomentum associated to a PND $ \mathbf{l} $ by \cref{eq:def-radiant-smomentum} is argued to vanish if there is no radiation propagating along the \emph{antipodal orientation} of the spatial projection to $ \scri $ of~$\mathbf{l}$, and indeed this happens if and only if $ \mathbf{l} $ is a repeated PND.

In the present work we have employed a much more explicit form of the general metric, leading to a clearer physical interpretation of the whole class of such black-hole spacetimes. Moreover, our main results, based on the evaluation of the asymptotic supermomentum and super-Poynting vectors, together with the new criterion for the presence of gravitational radiation, resolve a previous apparent  ``paradox'' in \cite{KrtousPodolsky:2005rd}, namely that the leading term \eqref{asyfield1+} of the Weyl tensor 
is \emph{nonzero even for} ${\alpha=0}$, that is also for \emph{non-accelerating} black holes. It is now clear that the presence/absence of gravitational radiation cannot be identified, in general, just from the leading Weyl term $\Psi_4^i$ (proportional to $\phi_2$) evaluated with respect to an interpretation tetrad along null geodesics arriving at $\scri$, but the existence of radiation actually depends
on the key parameter
$$
c= - \mathrm{i}\, \sinh\psi_s = - \mathrm{i}\, \tan \Theta_s \, ,
$$
see \eqref{radiation-condition} and \eqref{eq:sP-type-D}.  The deeper geometric and physical reason is that the new criterion is based on the more subtle \emph{quadratic} combination of the Weyl tensor (Bel-Robinson tensor, super-Poynting vector, superenergy density), not just on separate \emph{linear} components of the Weyl tensor. \\

In summary, we have demonstrated that
	\begin{equation}
		\cts{\P}{^a}=0\iff c=0 \iff \psi_{\rm s} =0  \iff \Theta_{\rm s} =0  \iff \alpha =0\ .
	\end{equation}

\section{Conclusions}
The recent covariant and gauge-invariant characterization of asymptotic gravitational radiation \cite{Fernandez-Senovilla:2022b} has been put to test by studying an improved form of the large family of type D  black holes \cite{Podolsky-Vratny:2023} for the case of a positive cosmological constant $ {\Lambda>0} $. In doing so, the metric has been conveniently adapted to investigate its conformal asymptotic structure, deriving the induced metric of the $\Lambda>0$ conformal boundary $ \scri $. \\

It has been shown that the characterisation in terms of the asymptotic super-Poynting vector field $ \cts{\P}{^a} $ --- that had been tested for other exact solutions \cite{Fernandez-Senovilla:2022b} already --- successfully accounts for the presence of gravitational radiation ($ \cts{\P}{^a}\neq 0 $) in this class of spacetimes, establishing that this is only possible if the black holes are accelerating ($ \alpha \neq 0 $). Furthermore, we have shown that the asymptotic super-Poynting vector is sensible to asymmetries of the pair of black holes' configurations, enhancing the relevance of this quantity to characterize and study the existence of radiation at infinity.\\

Thus, our study covers the characterization of gravitational radiation at $ \scri $ for a general class of type~D  (pairs of accelerating, charged, rotating, and NUTted)  black holes in the case with $ {\Lambda > 0}$. The limit to $ {\Lambda=0} $ can be taken appropriately \cite{Fernandez-Senovilla:2022a}. The next step is to investigate the radiative properties of more general black-hole exact solutions with $\Lambda$,   including recently found type~I black holes \cite{ChngMannStelea:2006gh,Podolsky:2020xkf,AstorinoBoldi:2023,Astorino:2023,Astorino:2024,Astorino:arXiv}, and also to consider the complementary case ${\Lambda<0}$. While this is left for a future work, we conjecture that the result will be consistent: only accelerating cases will have a non-vanishing asymptotic super-Poynting vector field.

\subsubsection*{Acknowledgements}
FFÁ is supported by the Grant Margarita Salas MARSA22/20 (Spanish Ministry of Universities and European Union), financed by European Union -- Next Generation EU. JP is supported by the Czech Science Foundation Grant No.~GA\v{C}R 23-05914S. JMMS is supported by Basque Government grant IT1628-22, and by Grant PID2021-123226NB-I00 funded by the Spanish MCIN/AEI/10.13039/ 501100011033 together with ``ERDF A way of making Europe''.

\section*{Appendices}
\appendix
\section{Details on the derivation of the new metric}\label{app:derivation-metric}

The recent improved form of the metric representing the family of type~D black holes, investigated here, which was presented in \cite{Podolsky-Vratny:2021, Podolsky-Vratny:2023}, reads
\begin{align}
\dd s^2 = \frac{1}{\Omega^2} &
  \bigg(\!-\frac{Q}{\rho^2}\big[\,\dd t- \big(a\sin^2\theta +4l\sin^2\!\tfrac{1}{2}\theta \big)\dd\varphi \big]^2
   + \frac{\rho^2}{Q}\,\dd r^2  \nonumber\\
& \quad  + \,\frac{\rho^2}{P}\,\dd\theta^2
  + \frac{P}{\rho^2}\,\sin^2\theta\, \big[ a\,\dd t -\big(r^2+(a+l)^2\big)\,\dd\varphi \big]^2
 \bigg), \label{newmetricGP2005}
\end{align}
where
\begin{eqnarray}
\Omega(r,\theta)    \rovno 1-\frac{\alpha\,a}{a^2+l^2}\, r\,(l+a \cos \theta) \,, \label{newOmega}\\[2mm]
\rho^2(r,\theta)    \rovno r^2+(l+a \cos \theta)^2 \,, \label{newrho}\\[2mm]
P(\theta) \rovno 1
   -2\,\Big(\,\frac{\alpha\,a}{a^2+l^2}\,\,m - \frac{\Lambda}{3}\,l \Big)(l+a\,\cos\theta)\nonumber\\
      &&\hspace{-2mm}  +\Big(\frac{\alpha^2 a^2}{(a^2+l^2)^2} (a^2-l^2 + e^2 + g^2) + \frac{\Lambda}{3} \Big)(l+a\,\cos\theta)^2 \,,
 \label{finalPagain}\\[2mm]
Q(r) \rovno \Big[\,r^2 - 2m\, r  + (a^2-l^2+e^2+g^2) \Big]
            \Big(1+\alpha\,a\,\frac{a-l}{a^2+l^2}\, r\Big)
            \Big(1-\alpha\,a\,\frac{a+l}{a^2+l^2}\, r\Big)\nonumber\\
   &&\hspace{-2mm}  - \frac{\Lambda}{3}\,r^2 \Big[\,r^2 + 2\alpha\,a\,l\,\frac{a^2-l^2}{a^2+l^2}\,r + (a^2+3l^2)\,\Big]\ , \label{finalQagain}
\end{eqnarray}
and the seven physical parameters $ m $, $ a $, $ l $, $ e $, $ g $, $ \alpha $, $ \Lambda $ are described in section~\ref{sec:newformmetric}.\\

When ${\Lambda=0}$, \emph{both} the metric functions $P$ and $Q$ are factorized \cite{Podolsky-Vratny:2021}. With ${\Lambda\not=0}$, it is possible to explicitly factorize the function $P$, and to compactify the function $Q$ as
\begin{eqnarray}
P(\theta) \rovno \Big(1 -\frac{\alpha\,a}{a^2+l^2}\,r_{\Lambda+}\,(l+a\,\cos\theta) \Big)
            \Big(1 -\frac{\alpha\,a}{a^2+l^2}\,r_{\Lambda-}\,(l+a\,\cos\theta) \Big) , \label{newP}\\[1mm]
Q(r) \rovno \big(r-r_{\Lambda+} \big) \big( r-r_{\Lambda-} \big)
            \Big(1+\alpha\,a\,\frac{a-l}{a^2+l^2}\, r\Big)
            \Big(1-\alpha\,a\,\frac{a+l}{a^2+l^2}\, r\Big) - \frac{\Lambda}{3}\,\Big[\,r^4 + \frac{(a^2+l^2)^2}{\alpha^2 a^2}\,\Big],\qquad \label{newQ}
\end{eqnarray}
using the two specific constants
\begin{eqnarray}
r_{\Lambda \pm} \equi  \mu \pm \sqrt{\mu^2 + l^2 - a^2 - e^2 - g^2 - \lambda}\,, \label{r+}
\end{eqnarray}
where
\begin{align}
\mu     \defeq    m - \frac{\Lambda}{3}\,l\,\frac{a^2+l^2}{\alpha\,a}\,,  \qquad
\lambda \defeq  \frac{\Lambda}{3}\,\frac{(a^2+l^2)^2}{\alpha^2 a^2} \,. \label{def-mu-and-lambda}
\end{align}
This is achievable provided ${\mu^2 + l^2 > a^2 + e^2 + g^2 + \lambda}$, in which case the expressions \eqref{r+} yield two distinct real constants (or a double root of $P$ given by ${r_{\Lambda+}  = r_{\Lambda-}= \mu}$ in the specific situation when ${\mu^2 + l^2 = a^2 + e^2 + g^2 + \lambda}$).\\

The general spacetime (\ref{newmetricGP2005}) contains \emph{electromagnetic field} represented by the Maxwell tensor $F_{ab}$, forming ${\bF = \tfrac{1}{2}F_{ab}\, \dd x^a \wedge \dd x^b =\dd\bA}$. Its 1-form potential ${\bA = A_a \dd x^a}$~is
\begin{equation}
 \bA =  - \frac{e\,r + g\,(l+a \cos \theta)}{r^2+(l+a \cos \theta)^2}\, \dd t
 + \frac{(e\,r + g\,l)\,(a \sin ^2 \theta + 4 l \sin ^2 \tfrac{1}{2}\theta)
+ g\,\big(r^2+(a+l)^2\big)\cos \theta}{r^2+(l+a \cos \theta)^2} \,\dd\varphi\,.
\label{vector-potential}
\end{equation}
Non-zero components of ${F_{ab} = A_{b,a}-A_{a,b}}$ are thus
\begin{align}
F_{rt}      &=\rho^{-4}\,\Big[\, e\,\big( r^2-(l+a \cos \theta)^2 \big) + 2\,g\,r\,(l+a \cos \theta) \Big]\,, \nonumber \\
F_{\varphi \theta} &= \rho^{-4}\,\Big[\, g\,\big( r^2-(l+a \cos \theta)^2 \big) - 2\,e\,r\,(l+a \cos \theta) \Big] \big(r^2 + (a+l)^2\big)\sin \theta  \,, \label{Fab}\\
F_{\varphi r} &=\big( a \sin ^2 \theta + 4 l \sin ^2 \tfrac{1}{2}\theta \big)\, F_{rt} \,, \nonumber \\
F_{\theta t} &= \frac{a}{r^2 + (a+l)^2}\,F_{\varphi \theta} \,. \nonumber
\end{align}
The electromagnetic field thus vanishes if (and only if) ${e=0=g}$.\footnote{It should be emphasized that the correct complete version of the electromagnetic field potential (\ref{vector-potential}) including both the electric and magnetic charges was first presented in paper~\cite{AstorinoBoldi:2023} by Astorino and Boldi, see Eq.~(3.31) and more detailed explanation in Appendix~A therein. Explicit relation to (\ref{vector-potential}) is obtained by the identification ${p=g}$, ${x=\cos\theta}$, ${{\cal R}^2=\rho^2}$, ${\Delta_r=Q}$, ${\Delta_x=P \sin^2\theta}$, and also (\ref{Fab}) agrees with (A.8)--(A.11) in~\cite{AstorinoBoldi:2023} when the opposite convention ${F_{ab} \to - F_{ab}}$ is taken into account.}\\

To derive the new form \eqref{newmetricJP2023} of the metric, convenient for investigation of an asymptotic structure of this large family of black-hole spacetimes, first we relabel the metric functions ${\Omega, \rho, Q}$ to ${\Omega', \rho', Q'}$, and then perform a simple transformation from $r$ to its \emph{reciprocal} coordinate
\begin{equation}
  q = \frac{1}{r}\,,
  \label{q=1/r}
\end{equation}
so that the metric (\ref{newmetricGP2005}) becomes
\begin{align}
\dd s^2 = \frac{1}{\Omega'^2} &
  \bigg(\!-\frac{Q'}{\rho'^2}\big[\,\dd t- \big(a\sin^2\theta +4l\sin^2\!\tfrac{1}{2}\theta \big)\dd\varphi \big]^2
   + \frac{\rho'^2}{q^4 Q'}\,\dd q^2  \nonumber\\
& \quad  + \,\frac{\rho'^2}{P}\,\dd\theta^2
  + \frac{P}{q^4\rho'^2}\,\sin^2\theta\, \big[ a\,q^2\,\dd t -\big(1+(a+l)^2q^2\big)\,\dd\varphi \big]^2
 \bigg). \label{newmetric-q}
\end{align}
Next we introduce new metric functions by the following rescaling
\begin{equation}\label{rescaled-funtions}
\Omega \defeq    q\, \Omega' \,, \qquad
\rho^2 \defeq    q^2 \rho'^2 \,, \qquad
Q      \defeq    q^4 Q' \,.
\end{equation}
A straightforward calculation then leads to the metric
\begin{align}
\dd s^2 = \frac{1}{\Omega^2} &
  \bigg(\!-\frac{Q}{\rho^2}\big[\,\dd t- \big(a\sin^2\theta +4l\sin^2\!\tfrac{1}{2}\theta \big)\dd\varphi \big]^2
   + \frac{\rho^2}{Q}\,\dd q^2  \nonumber\\
& \quad  + \,\frac{\rho^2}{P}\,\dd\theta^2
  + \frac{P}{\rho^2}\,\sin^2\theta\, \big[ a\,q^2\,\dd t -\big(1+(a+l)^2q^2\big)\,\dd\varphi \big]^2
 \bigg),
\end{align}
with the functions \eqref{newOmega-q}--\eqref{newQ-q}.

\subsection{Particular subcases of the metric on the conformal boundary~$\scri$}\label{A1}

The \emph{particular subcases}  of the metric $h$ given by eq.~\eqref{conformal-metric-final}, equivalent to \eqref{conformal-metric-final-alternative} or \eqref{conformal-metric-final-alternative-1}, are:

\begin{itemize}

\item For ${\alpha=0}$: {\bf Kerr-Newman-NUT-(A)dS} black hole without acceleration
$${\B=0\,, \qquad \C^2=1\,.}$$

\item For ${a=0}$: {\bf charged-NUT-(A)dS} without acceleration
$${\B=0\,, \qquad \C^2=1\,.}$$

\item For ${l=0}$: {\bf accelerating Kerr-Newman-(A)dS} black hole
$${\B=\alpha\,, \qquad \C^2=1 + \alpha^2 a^2\,.}$$

\item For ${\alpha=0=l}$: {\bf Kerr-Newman-(A)dS} black hole
$${\B=0\,, \qquad \C^2=1\,.}$$

\end{itemize}

\vspace{2mm}

An important special case is the class of {\bf Kerr-(A)dS} black holes (${\alpha, l, e, g =0}$), for which ${q_\scrisub=0}$, ${\rho^2_\scrisub =1}$,  ${Q_\scrisub   = -\frac{\Lambda}{3}}$, ${P = 1 + \frac{\Lambda}{3}\,a^2\cos^2\theta}$, so that the metric \eqref{conformal-metric-final-alternative} reduces to
\begin{align}
h =
\frac{\Lambda}{3}\,\dd t^2
+ \Big( 1 + \frac{\Lambda}{3}\,a^2\Big)\sin^2\theta\,\dd\varphi^2
- 2\,\frac{\Lambda}{3}\,a\sin^2\theta\,\dd t\, \dd\varphi
+ \Big( 1 + \frac{\Lambda}{3}\,a^2\cos^2\theta  \Big)^{-1}\dd\theta^2\,.
\end{align}

Another prominent metric is the {\bf C-metric with $\Lambda$} (${a, l, e, g =0}$). In such a case
${q_\scrisub=\alpha\cos\theta}$, ${\rho^2_\scrisub =1}$, ${Q_\scrisub = -\frac{\Lambda}{3} - \alpha^2 P \sin^2\theta}$, ${P = 1 - 2\alpha m \cos\theta}$, and \eqref{conformal-metric-final-alternative} thus simplifies to
\begin{align}
h = \Big( \frac{\Lambda}{3} + \alpha^2 P \sin^2\theta \Big) \dd t^2 + P \sin^2\theta\,\dd\varphi^2
+\frac{\Lambda}{3}\Big(\frac{\Lambda}{3} + \alpha^2 P \sin^2\theta \Big)^{-1}\,\frac{\dd\theta^2}{P} \,.
    \label{C-AdS-metric}
\end{align}

Without acceleration ${\alpha=0}$, we get the {\bf Schwarzschild-(A)dS} metric on $\scri$, namely
\begin{align}
h = \frac{\Lambda}{3}\,\dd t^2 +\dd\theta^2 + \sin^2\theta\,\dd\varphi^2 \, .
    \label{Schw-AdS-metric}
\end{align}

\section{General considerations for type D spacetimes}\label{app:general-type-D}
	For spacetimes with algebraically special, double-degenerate type~D rescaled Weyl tensor $ \ct{d}{_{\alpha\beta\gamma}^\delta} $ at~$\scri$ --- as it is the case of the metric \eqref{conformal-metricJP2023} ---, it is possible to prescribe a \emph{general construction of strongly oriented tetrads}. Let $ \mathbf{n} $ be the unit (${\mathbf{n}\cdot\mathbf{n} = -1}$), future-pointing normal to $\scri$,   cf.~\eqref{eq:unit-normal-scri},\footnote{Which is well defined at $ \scri $ and in its neighbourhood.} and consider the tetrad $ {(\textbf{k},\textbf{l},\textbf{m})} $, where $\textbf{k}$ and $\textbf{l}$ are future-oriented and aligned with the two double PNDs. Next, we define a real and a complex scalar, $ b $ and $ c $, as
		\begin{align}
			b^2&\defeq -\frac{1}{\sqrt{2}}\,\frac{1}{\ct{k}{_{\alpha}}\ct{n}{^\alpha}}\ ,\label{eq:boost-paramenter-b}\\
			c  & \defeq \sqrt{2}\,\,\ct{\bar{m}}{_\alpha}\ct{n}{^\alpha}.\label{eq:rotation-parameter-c}
		\end{align}
	With the scalar~$b$, let us perform a boost of the null directions to an auxiliary tetrad
		\begin{align}
		\cta{\textbf{k}}{}&=b^2\,\ct{\textbf{k}}{}\,, \nonumber\\
		\cta{\textbf{l}}{}&=\frac{1}{b^2}\,\ct{\textbf{l}}{}\,,\\
		\cta{\textbf{m}}{}&=\ct{\textbf{m}}{}\,, \nonumber
		\end{align}
	and subsequently, with the complex scalar~$c$, apply a null rotation around $ \cta{\textbf{k}}{} $,
		\begin{align}
		 \ctst{\textbf{k}}{}&=\cta{\textbf{k}}{}\,, \nonumber\\
		 \ctst{\textbf{l}}{}&=\cta{\textbf{l}}{}+c\,\cta{\textbf{m}}{}+\bar{c}\,\cta{\bar{\textbf{m}}}{}+c\bar{c}\,\cta{\textbf{k}}{}\,,\\
		 \ctst{\textbf{m}}{}&=\cta{\textbf{m}}{}+\bar{c}\,\cta{\textbf{k}}{}\,. \nonumber
		\end{align}
	Since
		\begin{equation}\label{eq:coplanarity}
			\textbf{n} = \frac{1}{\sqrt{2}} \prn{ \ctst{\textbf{k}}{} + \ctst{\textbf{l}}{}}\,,
		\end{equation}
	at $\scri$ the strong orientation is given by the space-like unit vector field
		\begin{equation}
			\textbf{M} \defeqs \frac{1}{\sqrt{2}}\prn{\,\ctst{\textbf{l}}{} - \ctst{\textbf{k}}{}}= \ct{\textbf{n}}{}-\sqrt{2}\,  \ctst{\textbf{k}}{}\ ,\qquad \textbf{M}\cdot\textbf{M}=1\ ,\quad \textbf{n}\cdot\textbf{M}=0\,.
		\end{equation}
	From here, one can already make an observation about the content of gravitational radiation at $ \scri $. As  follows from results of \cite{Fernandez-Senovilla:2022b}, for a type~D rescaled Weyl tensor at $ \scri $,  the gravitational radiation condition is equivalent to the \emph{coplanarity of the normal to~$\scri$ and the two PNDs}. Hence,
		\begin{equation}\label{radiation-condition}
		\cts{\P}{^a}=0 \quad\iff\quad \text{no gravitational radiation at }\scri \quad\iff\quad c=0 \,.
		\end{equation}
		This shows that the \emph{complex scalar} $ c $ of \cref{eq:rotation-parameter-c} --- or, equivalently, the components of the normal $ \textbf{n}$ projected to the two-dimensional plane orthogonal to the PNDs --- \emph{encodes the information about gravitational radiation}.\\
		
		 To see more explicitly the role played by $ c $ in the radiation condition, one can write the \emph{Cotton-York tensor} --- which up to a constant factor coincides with the magnetic part $ \ct{C}{_{\alpha\beta}} $ of the rescaled Weyl tensor of $ \ct{d}{_{\alpha\beta\gamma}^\delta} $ --- and the electric part $ \ct{D}{_{\alpha\beta}} $. First observe that in the `natural' tetrad $ {(\textbf{k},\textbf{l},\textbf{m})} $ the only non-vanishing Weyl scalar of $ \ct{d}{_{\alpha\beta\gamma}^\delta} $ is $ \phi_{2} $. Then, in the strongly oriented tetrad $ {(\ctst{\textbf{k}}{},\ctst{\textbf{l}}{},\ctst{\textbf{m}}{})} $, the non-vanishing scalars are:\footnote{The boost $ b $ does not affect $ \phi_{2} $, whereas the null rotation $ c $ produces the rest of the scalars.}
		 	\begin{align}
		 		\ctst{\phi}{_{4}}&=6c^2\,\ct{\phi}{_{2}}\,, \nonumber\\	
		 		\ctst{\phi}{_{3}}&=3c  \,\ct{\phi}{_{2}}\,,\label{eq:sphi3}\\	 	 	
	 			\ctst{\phi}{_{2}}&=    \ct{\phi}{_{2}}\,. \nonumber
		 	\end{align}
		 Following the formulae of section 2 and appendix D of \cite{Fernandez-Senovilla:2022b}, we can write\footnote{Latin indices could be used instead, as the fields involved in these expressions are all completely tangent to $ \scri $.}
		 	\begin{align}
		 		 \ct{D}{_{\alpha\beta}}&=\Re\prn{\ct{\phi}{_{2}}}\,(3\ct{M}{_{\alpha}}\ct{M}{_{\beta}}-\ms{_{\alpha\beta}})+\Big[-3\sqrt{2}\,c\,\ct{\phi}{_{2}}
\,\ct{M}{_{(\alpha}}\ctst{m}{_{\beta)}}-3\sqrt{2}\,\bar{c}\,\ct{\bar{\phi}}{_{2}}\,\ct{M}{_{(\alpha}}\ctst{\bar{m}}{_{\beta)}}\nonumber\\
		 		 &\hspace{60mm}
+3c^2\ct{\phi}{_{2}}\,\ctst{m}{_{\alpha}}\,\ctst{m}{_{\beta}}
+3\bar{c}^2\ct{\bar{\phi}}{_{2}}\,\ctst{\bar{m}}{_{\alpha}}\,\ctst{\bar{m}}{_{\beta}}\Big]\, ,\label{eq:electric-part-type-D}\\ \ct{C}{_{\alpha\beta}}&=\Im\prn{\ct{\phi}{_{2}}}\,(3\ct{M}{_{\alpha}}\ct{M}{_{\beta}}-\ms{_{\alpha\beta}})+\mathrm{i}\,
\Big[-3\sqrt{2}\,c\,\ct{\phi}{_{2}}\,\ct{M}{_{(\alpha}}\ctst{m}{_{\beta)}}+3\sqrt{2}\,\bar{c}\,\ctst{\bar{\phi}}{_{2}}\,\ct{M}{_{(\alpha}}\ctst{\bar{m}}{_{\beta)}}\nonumber
		 		\\
		 		&\hspace{60mm}
+3c^2\ct{\phi}{_{2}}\,\ctst{m}{_{\alpha}}\,\ctst{m}{_{\beta}}
-3\bar{c}^2\ct{\bar{\phi}}{_{2}}\,\ctst{\bar{m}}{_{\alpha}}\ctst{\bar{m}}{_{\beta}}\Big]\,.\label{eq:magnetic-part-type-D}
		 	\end{align}
		 Notice that  $ \prn{\textbf{M}, \ctst{\textbf{m}}{}, \ctst{\bar{\textbf{m}}}{}}$ is a basis on $ \scri $. A couple of results from \cite{Fernandez-Senovilla:2022b} implies that for type~D spacetimes there is no gravitational radiation if and only if $ {\ct{D}{_{\alpha\beta}}} $ and $ \ct{C}{_{\alpha\beta}} $ are proportional to $ {(3\ct{M}{_{\alpha}}\ct{M}{_{\beta}}-\ms{_{\alpha\beta}})} $. From \cref{eq:electric-part-type-D,eq:magnetic-part-type-D} we thus conclude that this occurs if and only if $ {c=0} $ --- excluding the conformally flat type~O case. Thus, whenever $ {c=0} $ the electric and magnetic parts of the rescaled Weyl tensor are proportional, implying the vanishing of their matrix commutator --- that is the vanishing of the asymptotic super-Poynting vector $ \cts{\P}{^a} $. \\

Observe that  $ \ct{\phi}{_{2}} $ \emph{on its own does not contain} the information about gravitational radiation at~$\scri$. It is possible to have
no gravitational radiation  with $ {\ct{\phi}{_{2}}\neq 0} $ at $ \scri $ if $ {c=0} $. The case $ {\phi_{2}=0} $ is trivial and corresponds to vanishing $ \ct{d}{_{\alpha\beta\gamma}^\delta} $.\\
		
		 Indeed, the \emph{asymptotic super-Poynting vector field} reads
		 	\begin{equation}\label{eq:sP-type-D}
	 \cts{\P}{^a}=-9\sqrt{2}\,\ct{\phi}{_{2}}\ct{\bar{\phi}}{_{2}}\,
          (1+2\,c\bar{c}) \brkt{c\,\ctst{m}{^a}+\bar{c}\,\ctst{\bar{m}}{^a}+\sqrt{2}\,c\bar{c}\,\ct{M}{^a}}.
		 	\end{equation}
		 It is thus clear that, for $ {\ct{\phi}{_{2}}\neq 0} $ (i.e., $ {\ct{d}{_{\alpha\beta\gamma}^\delta}\neq 0} $), $ {\cts{\P}{^{\,a}}=0} $ iff $ {c=0} $. Hence, in agreement with \cite{Fernandez-Senovilla:2022b}, the superenergy cannot propagate in directions orthogonal to $ \ct{M}{^a} $. Observe also that
		 	\begin{equation}
		 		\ct{M}{_{a}}\cts{\P}{^{\,a}}\leq 0\,.
		 	\end{equation} 	

		 The expression for the  \emph{asymptotic super-energy density}  is
		 	\begin{equation}
		 		\W=6\,\ct{\phi}{_{2}}\ct{\bar{\phi}}{_2}\,\big( 1+6\,c\bar{c}+6\,c^2\bar{c}^2 \big)\,,
		 	\end{equation}
		 which can be different from zero for $ {c=0} $.  In particular, for the Schwarzschild-de~Sitter black hole (the solution by Kottler), we get $ {\W=6m^2} $. This is completely fine, as the gravitational \emph{radiation} is determined by the super-Poynting vector. \\

	 	One could also choose \emph{the other strong orientation}   complementary to \eqref{eq:boost-paramenter-b}, by defining
	 		\begin{equation}\label{eq:boost-parameter-other-strong-orientation}
	 		\tilde{b}^2 \defeq -\frac{1}{\sqrt{2}\,\,\ct{l}{_{\alpha}}\ct{n}{^\alpha}}\ ,
	 		\end{equation}
	 	and keeping the same $ c $. Thus, the tetrad adapted to such strong orientation reads
		 	\begin{align}
			 	 \ctst{\tilde{\textbf{k}}}{}&=\cta{\tilde{\textbf{k}}}{}+c\,\cta{\textbf{m}}{}
                    +\bar{c}\,\cta{{\bar{\textbf{m}}}}{}+c\,\bar{c}\,\cta{\tilde{\textbf{l}}}{}\,, \nonumber\\
			 	\ctst{\tilde{\textbf{l}}}{}&=\cta{\tilde{\textbf{l}}}{} \,,\\
			 	 \ctst{\tilde{\textbf{m}}}{}&=\cta{\tilde{\textbf{m}}}{}+\bar{c}\,\cta{\tilde{\textbf{l}}}{}\,, \nonumber
			\end{align}
		where now
			\begin{align}
			\cta{\tilde{\textbf{k}}}{}&=\frac{1}{\tilde{b}^2}\,\ct{\textbf{k}}{}\,, \nonumber\\
			\cta{\tilde{\textbf{l}}}{}&=\tilde{b^2}\,\ct{\textbf{l}}{}\,,\\
			\cta{\textbf{m}}{}&=\ct{\textbf{m}}{}\,. \nonumber
			\end{align}
		One defines the new vector field  tangent to $ \scri $ as

		\begin{equation}
			\tilde{\textbf{M}} \defeqs \frac{1}{\sqrt{2}}\prn{\,\ctst{\tilde{\textbf{l}}}{} - \ctst{\tilde{\textbf{k}}}{}}= \ct{\textbf{n}}{}-\sqrt{2}\,  \ctst{\tilde{\textbf{k}}}{}\ ,\qquad \tilde{\textbf{M}}\cdot\tilde{\textbf{M}}=1\ ,\quad \textbf{n}\cdot\tilde{\textbf{M}}=0\,.
		\end{equation}

%
		Using  \cref{eq:coplanarity} and its version with tildes, it follows that
			\begin{equation}\label{eq:relation-c-b}
				c\bar{c}=\frac{\gamma\,(1-2\gamma)+1}{\gamma\prn{1+2\gamma}}
			\end{equation}
		where the coefficient
			\begin{equation}
				\gamma\defeq (b\tilde{b})^2\ \in (0,1]
			\end{equation}
		has been introduced. The range follows from the  normalization condition on the lightlike PNDs, namely ${\textbf{k}\cdot\textbf{l}=-1}$. With \cref{eq:relation-c-b}, it is possible to write
			\begin{equation}
				\textbf{M}\cdot\tilde{\textbf{M}} = 2\gamma-1\,.
			\end{equation}
	Notice that $ {\gamma=1} $ is the only root of the numerator in \cref{eq:relation-c-b} allowed by the range of $ \gamma $, and thus
			\begin{equation}
				\textbf{M}\cdot\tilde{\textbf{M}} = 1 \quad\iff\quad \gamma=1 \quad\iff\quad \cts{\P}{^{\,a}}=0\,,
			\end{equation}
		as expected.

 		\subsection{Radiant superenergy}\label{sec:radiant-se-general}
	 	Using the same definitions from above, the radiant superenergy associated with the lightlike directions read, respectively:
		 	\begin{align}
		 		\ctl{\Z}{}&=\ctkt{\Z}{}=36\,c\bar{c}\,\ct{\phi}{_{2}}\bar{\ct{\phi}{_{2}}}\,,\\
		 		\ctl{\W}{}&=\ctkt{\W}{}=144\,c^2\bar{c}^2\,\ct{\phi}{_{2}}\bar{\ct{\phi}{_{2}}}\,,\\
		 	    \ctlt{\Q}{^\alpha}&=0=\ctk{\Q}{^\alpha}\ ,\\
		 		 \ctkt{\Q}{^\alpha}&=36\,c\bar{c}\,\ct{\phi}{_{2}}\bar{\ct{\phi}{_{2}}}\prn{4c\bar{c}\,\ctst{\tilde{l}}{^\alpha}+\ctst{\tilde{k}}{^\alpha}
-2c\,\ctst{\tilde{m}}{^\alpha}-2\bar{c}\,\ctst{\bar{\tilde{m}}}{^\alpha}},\\
		 		 \ctl{\Q}{^\alpha}&=36\,c\bar{c}\,\ct{\phi}{_{2}}\bar{\ct{\phi}{_{2}}}\prn{4c\bar{c}\,\ctst{k}{^\alpha}+\ctst{l}{^\alpha}
-2c\,\ctst{{m}}{^\alpha}-2\bar{c}\,\ctst{\bar{{m}}}{^\alpha}}.
		 	\end{align}
		 It can be easily verified that the radiant supermomenta $ \ctl{\Q}{^\alpha} $ and $ \ctkt{\Q}{^\alpha} $ are lightlike. Their tangent part to $ \scri $, denoted here with an overline, can be expressed as
		 	\begin{align}
			 \ctkt{\overline{\Q}}{^\alpha}&=18\sqrt{2}\,c\bar{c}\,\ct{\phi}{_{2}}\bar{\ct{\phi}{_{2}}}
\brkt{\prn{1-4c\bar{c}}\ct{\tilde{M}}{^\alpha}-2\sqrt{2}\,c\,\ctst{\tilde{m}}{^\alpha}-2\sqrt{2}\,\bar{c}\,\ctst{\bar{\tilde{m}}}{^\alpha}},\\
		 		 \ctl{\overline{\Q}}{^\alpha}&=18\sqrt{2}\,c\,\bar{c}\,\ct{\phi}{_{2}}\bar{\ct{\phi}{_{2}}}
\brkt{\prn{1-4c\bar{c}}\ct{M}{^\alpha}-2\sqrt{2}\,c\,\ctst{{m}}{^\alpha}-2\sqrt{2}\,\bar{c}\,\ctst{\bar{{m}}}{^\alpha}}.
		 	\end{align}	
		 It is a matter of direct calculation using the relations between the different tetrads to arrive at
		 	\begin{equation}
		 		\ctl{\Q}{^\mu}\,\ctkt{\Q}{_{\mu}}=-(36\,c\bar{c}\,\ct{\phi}{_{2}}\ct{\bar{\phi}}{_{2}})^2\,\gamma\, \leq 0\ ,
		 	\end{equation}
		 equality holding if and only if there is no gravitational radiation. Also,
		 	\begin{equation}
		 		T\defeq \ct{n}{_{\mu}}\ctkt{\Q}{^\mu}=\ct{n}{_{\mu}}\ctl{\Q}{^\mu}=36\sqrt{2}\,c\bar{c}\,\phi_{2}\bar{\phi}_{2}\prn{\frac{3}{2}-\frac{2}{\gamma}} ,
		 	\end{equation}
		 and then
		 	\begin{equation}
		 		 \ctkt{\overline{\Q}}{^\mu}\,\ctkt{\overline{\Q}}{_{\mu}}=\ctl{\overline{\Q}}{^\mu}\,\ctl{\overline{\Q}}{_{\mu}}=T^2\,.
		 	\end{equation}
		 One can write
		 	\begin{align}
		 		\ctl{\Q}{^\alpha}&=-T\prn{N^\alpha+\ctl{\overline{q}}{^\alpha}}\ ,\quad \ctl{\overline{q}}{^\mu}\,\ctl{\overline{q}}{_{\mu}}=1\ ,\quad \ctl{\overline{q}}{^\mu}\,\ctl{N}{_{\mu}}=0\,,\\
		 		\ctkt{\Q}{^\alpha}&=-T\prn{N^\alpha+\ctkt{\overline{q}}{^\alpha}}\ ,\quad \ctkt{\overline{q}}{^\mu}\,\ctkt{\overline{q}}{_{\mu}}=1\ ,\quad \ctkt{\overline{q}}{^\mu}\,\ctl{N}{_{\mu}}=0\,,
		 	\end{align}
		 where the unit vector fields $ \ctl{\mathbf{\overline{q}}}{} $ and $ \ctkt{\mathbf{\overline{q}}}{} $ give the spatial component of propagation of the `tidal momentum' as measured by $ \ctst{\mathbf{l}}{} $ and $ \tilde{\ctst{\mathbf{k}}{}} $, respectively. To characterise  their relative orientation, it is useful to introduce the coefficient
		 	\begin{equation}\label{eq:beta}
		 		 \beta \defeq \ctl{\overline{q}}{^\mu}\,\ctkt{\overline{q}}{_{\mu}}=1-\frac{2\gamma^3}{\prn{3\gamma-4}^2}\,.
		 	\end{equation}
		 Observe that $ \beta\in[-1,1) $. The condition $ \beta=0 $ gives only one real root, namely ${\gamma=0.918 \ldots}$.

\printbibliography

\end{document}